\documentclass[12pt]{article}
\usepackage{amssymb}
\usepackage{amsfonts}
\usepackage{mathrsfs}
\usepackage{latexsym}
\usepackage{amsmath}
\usepackage{graphics}
\usepackage{graphicx}
\usepackage{sectsty}
\usepackage{setspace}
\usepackage{multirow}
\sectionfont{\normalsize\bf}
\subsectionfont{\rm\normalsize}
\def\be{\begin{equation}}
\def\ee{\end{equation}}
\def\bal{\begin{align}}
\def\ealn{\end{align}}

\newcommand{\bpi}{\bar{\pi}}

\input amssym.def
\input amssym
\def\Rop{{\Bbb R}}

\def\bvac{\langle 0|}

\def\R{{\cal R}}
\def\A{{\cal A}}
\def\O{{\cal O}}
\def\N{{\cal N}}
\def\P{{\cal P}}
\def\C{{\cal C}}
\def\M{{\cal M}}
\def\ba{{\overline a}}
\def\bc{{\overline c}}
\def\bpi{{\overline\pi}}
\def\bbeta{{\overline \beta}}
\def\bV{{\overline V}}
\def\bw{{\overline w}}
\def\bv{{\overline v}}
\def\bW{{\overline W}}
\def\bA{{\bar A}}

\def\vs{\vskip6pt}
\def\vss{\vskip12pt}

\numberwithin{equation}{section}

\begin{document}

\thispagestyle{empty}
\begin{flushright}
\end{flushright}
\baselineskip=16pt
\vspace{.5in}
{
\begin{center}
{\bf Gluon Tree Amplitudes in Open Twistor String 
Theory}
\end{center}}
\vskip 1.1cm
\begin{center}
{Louise Dolan}
\vskip5pt

\centerline{\em Department of Physics}
\centerline{\em University of North Carolina, Chapel Hill, NC 27599} 
\bigskip
\bigskip        
{Peter Goddard}
\vskip5pt

\centerline{\em Institute for Advanced Study}
\centerline{\em Princeton, NJ 08540, USA}
\bigskip
\bigskip
\bigskip
\bigskip
\end{center}

\abstract{\noindent 
We show how the link variables of Arkani-Hamed, Cachazo, Cheung and Kaplan 
(ACCK),
can be used to compute general gluon tree amplitudes in the twistor string. 
They arise from instanton sectors labelled by $d$, with $d=n-1$,
where $n$ is the number of negative helicities. Read backwards, 
 this shows how the various forms for the tree amplitudes studied 
by ACCK can be grouped into contour integrals whose structure implies the 
existence of an underlying string theory.
\bigskip

\setlength{\parindent}{0pt}
\setlength{\parskip}{6pt}

\setstretch{1.05}
\vfill\eject
\vskip50pt

\section{\bf Introduction}

Inspired by work of Arkani-Hamed, Cachazo, Cheung and Kaplan 
\cite{ACCK1,ACCK2}, we 
use link variables to
obtain expressions for tree amplitudes in open  twistor string theory \cite{B}-\cite{W}.
This extends the derivation from the twistor string 
beyond maximally helicity violating
(MHV) amplitudes  and special cases of non-MHV trees \cite{BerkMotl}-\cite{CSW},
as well as providing a basis for the dual structure
envisioned in ACCK. We use a canonical quantization \cite{DG1,DG2}
of Berkovits' version of twistor string theory, and compute
the gluon trees.

The extensive literature for amplitudes in the spinor heliciy basis,
sampled by \cite{PT}-\cite{BDDK}, has been used in developing 
recursion relations \cite{BCF,BCFW}. These were motivated
by a remarkable formulation of string theory on twistor space \cite{W},
which made contact with a twistor description for gauge theory \cite{P,N}.
Additional dual forms for trees are found \cite{H1}-\cite{DH}.

At loop level, the twistor string has been difficult to interpret 
as a dual for the gauge theory \, \cite{BW, DG1}. In this
paper, however,  we show that the gauge theory based analysis of ACCK, 
which is phrased in terms of link variables, appears naturally to lead back
to the twistor string at tree level. This may eventually enlighten our 
treatment of string loops, and the pursuit of the dual S-matrix.  

Suppose we have $N$ gluons, labeled $\alpha = 1,\ldots  N$, 
with momenta ${p^a_\alpha}_{\dot a}
=\pi_\alpha^a\bpi_{\alpha\dot a},$ and helicities $\epsilon_\alpha$, 
$m$ of which are positive and $n$ negative,
$m+n=N$.
Write $\P$ for the set of positive 
helicity particles and $\N$ for the set of negative helicity particles. 

The link variables $c_{ir},i \in\P, r\in\N$ satisfy the $2N$ linear equations 
\begin{align}
\pi_j &=\sum_{r\in\N} c_{jr}\pi_r\label{L1}\\
\bar\pi_s &=-\sum_{i\in\P}\bar\pi_i c_{is}.\label{L2}
\end{align}
where we have suppressed the spinor indices.
See \cite{DG1} for our conventions.
These equations are not independent because, as noted by ACCK, 
they imply momentum conservation.
As in \cite{ACCK1} (Eq.37), these linear conditions 
imply energy-momentum conservation:
\begin{align}
\sum_{j\in\P}\bar\pi_{j\dot a}\pi_j^a&=\sum_{j\in\P}\sum_{r\in\N}\bar\pi_{j\dot a}c_{jr}\pi_r^a=-\sum_{r\in\N}\bar\pi_{r\dot a}\pi_r^a;
\label{EMC}\end{align}
for momenta satisfying this consistency condition they provide $N'=2N-4$ 
constraints on the $mn$ variables $c_{ir}$, leaving 
\be N_R=mn-N'=(m-2)(n-2)\label{NR}\ee
degrees of freedom. The philosophy outlined by ACCK is to seek to write the 
tree and loop amplitudes of gauge theory  as contour integrals over the 
remaining $N_R$ degrees of freedom. 
 
The main observation underlying the analysis of this paper is that, 
in open twistor string theory, 
the link variables should be of the form 
\be c_{js}={k_j\over k_s(\rho_j-\rho_s)},\label{cjs}\ee
for some suitable $\rho_\alpha, k_\alpha$, as we shall see at the beginning 
of section 3. 
The necessary and sufficient condition 
for the link variables to be of this form is that the matrix 
\begin{align}
\left( \begin{matrix}
(c_{i_1r_1})^{-1}&(c_{i_1r_2})^{-1}&\ldots &(c_{i_1r_n})^{-1}\\
(c_{i_2r_1})^{-1}&(c_{i_2r_2})^{-1}&\ldots &(c_{i_2r_n})^{-1}\\
\vdots&\vdots& &\vdots\\
(c_{i_mr_1})^{-1}&(c_{i_mr_2})^{-1}&\ldots &(c_{i_mr_n})^{-1}
\end{matrix}\right)\label{C}\end{align}
should have rank two, as we discuss in more detail in section 5. This is equivalent 
to the vanishing of the determinant of each $3\times 3$ submatrix, that is of 
each determinant of the form
\begin{align}
\C^{ijk}_{rst}&=\left|\begin{matrix} c_{is}c_{it}&c_{it}c_{ir}& c_{ir}c_{is}\cr
c_{js}c_{jt}&c_{jt}c_{jr}& c_{jr}c_{js}\cr
c_{ks}c_{kt}&c_{kt}c_{kr}& c_{kr}c_{ks}\cr\end{matrix}\right|\cr
\noalign{\vskip3pt}
 &=c_{ir}c_{js}c_{kt}c_{jr}c_{ks}c_{it}+c_{jr}c_{ks}c_{it}c_{kr}c_{is}c_{jt}+c_{kr}c_{is}c_{jt}c_{ir}c_{js}c_{kt}\cr
&\qquad\qquad-c_{kr}c_{js}c_{it}c_{jr}c_{is}c_{kt} - 
c_{jr}c_{is}c_{kt}c_{ir}
c_{ks}c_{jt} -  c_{ir}c_{ks}c_{jt}c_{kr}c_{js}c_{it}.
\label{C'}\end{align}
For this condition to be met, it is sufficient for a suitable subset comprising $N_R$ of the $\C^{ijk}_{rst}$ to vanish, {\it e.g.} if we fix $I,J\in\P$ and $R,S\in\N$, it is sufficient to have the vanishing of the $N_R$, quantities $\C^{IJk}_{RSt}$ where $k$ ranges over $\P'$,  the remaining $m-2$ elements of $\P$, and $t$ ranges over $\N'$, the remaining $n-2$ elements of $\N$. 
Using the linear conditions (\ref{L1}), (\ref{L2}) to express the $c_{ir}$ 
in terms of the $c_{kt}, k\in\P', t\in\N'$, we find in section 3 that 
the tree amplitude will have the form
\be\oint F(c) \prod_{k\in\P'\atop t\in\N'} {dc_{kt}\over \C_{kt}},\label{ointF}\ee
where $F(c)$ is a simple rational function of the $c_{ir}$.

In section 2, we review the derivation of vertex operator 
expressions for the general $N$-point tree 
amplitudes in twistor string theory from vertex operators. 
In section 3, we analyze the amplitude as an integral over constraints.
In section 4, we derive the integrand function $F(c)$, 
as a function of the link variables, from twistor string theory.
In section 5, we discuss the parametrization of the linear constraints, and 
complete the description of the contour integral expression for the amplitudes.
In section 6, we compute all 6-point functions, including
 alternativeforms, by evaluating the contour integral as a 
sum over residues. In section 7, we use our general formulae to check 
the 7-pt tree with alternating helicities.

\section{\bf The $N$-point Amplitude}
\nobreak
As in \cite{B,DG1}, 
we consider  conjugate twistor variables $Z$ and $W$
\be
Z=\left(\begin{matrix}\pi^a\cr \omega^{\dot a}\end{matrix}\right),\qquad 
W=\left( \begin{matrix}
\bar\omega_a\cr\bar\pi_{\dot a}\cr \end{matrix}\right),\ee
\be
W\cdot Z =\bar\omega\cdot\pi + \bar\pi\cdot\omega\equiv\bar\omega_a\pi^a 
+ \bar\pi_{\dot a}\omega^{\dot a},
\ee
and the field describing the twistor string,
\be Z(\rho)=\left(\begin{matrix}\lambda^a(\rho)\cr \mu^{\dot a}(\rho)\end{matrix}
\right).\nonumber\ee
We fourier transform the open string vertex operators for gluons 
according to their helicity\cite{ACCK1}, as 
\be
V^A_+(W,\rho)=\int d^2\pi^a e^{i\bar\omega_a\pi^a}
\int {d\kappa\over \kappa}\delta^2(\kappa\lambda^a(\rho)-\pi^a)
e^{i\kappa\bar\pi_{\dot b}\mu^{\dot b}(\rho)}J^A
=\int {d\kappa\over \kappa}e^{i\kappa W\cdot Z(\rho)}J^A,\ee
\begin{align}
V^A_-(Z,\rho)&=\int d^2 \bar\pi_{\dot a}e^{-i\omega^{\dot a}\bar\pi_{\dot a}}
\int \kappa^3d\kappa\delta^2(\kappa\lambda^a(\rho)-\pi^a)
e^{i\kappa\bar\pi_{\dot b}\mu^{\dot b}(\rho)}J^A\psi^1\ldots\psi^4\cr
&=\int \kappa^3d\kappa \delta^4(\kappa Z(\rho)-Z)J^A\psi^1\ldots\psi^4.
\end{align}
 Defining 
$X_j=W_j,\; j\in\P;\quad X_s=Z_s,\; s\in\N,$ we
compute the tree amplitudes as a sum over instanton sectors. 
The only non-vanishing contribution to any tree with $n$ negative
helicity states is from the sector with instanton number $d$,
where $d=n-1$, \cite{DG1} 
\begin{align}
M^{\epsilon_1\ldots\epsilon_N}
&=\int \bvac V^{A_1}_{\epsilon_1}(X_1,\rho_1)\ldots V^{A_N}_{\epsilon_N}
(X_N,\rho_N) |0\rangle
\prod_{\alpha=1}^N d\rho_\alpha\Big/dg\cr
&=\int \prod_{\alpha=1}^N {d\rho_\alpha d\kappa_\alpha\over \kappa_\alpha} 
\bvac e^{(n-1)q_0}\prod_{ s \in\N}\delta^4(\kappa_s  Z(\rho_s )-Z_s )
\exp\left\{i\sum_{ j \in\P}\kappa_ j  W_ j \cdot Z(\rho_ j )\right\}
|0\rangle\cr
&\hskip20truemm\times \prod_{ s\in\N} \kappa_ s ^4 \prod_{r < s;r,s\in\N}
(\rho_r -\rho_s)^4\bvac J^{A_1}(\rho_1)J^{A_2}(\rho_2)\ldots J^{A_N}(\rho_N)
|0\rangle \Big/ dg\label{M1}\cr
\end{align}
and $dg$ is the invariant measure on the group $GL(2,\Rop)$ of M\"obius and scale transformations.
Because $Z(\rho)$ is a polynomial of order $n-1$,
\be Z(\rho) =\sum_{ s \in\N}{1\over \kappa_ s } Z_ s \prod_{{ r \ne  s ; r \in\N}}{\rho-\rho_ r \over \rho_ s -\rho_ r },\ee 
so that $\kappa_ r  Z(\rho_ r )=Z_ r $, $  r \in\N$.
If $\xi_ r = \kappa_ r  Z(\rho_ r )-Z_ r $,  $ r \in\N$, the Jacobian 
resulting from performing the integrations corresponding to the zero modes of 
$Z(\rho)$ is
$${\partial(\xi_ j : j \in\N)\over \partial(Z_{-m+1},\dots ,Z_0)}
=\prod_{s\in\N} \kappa_ s  \prod_{  r < s ;  r , s \in\N}(\rho_ r -\rho_ s ).$$
This factor will cancel the worldsheet fermion contribution included in 
(\ref{M1}), and the amplitude becomes 
\begin{align}
M^{\epsilon_1\ldots\epsilon_N}
&=\int \prod_{\alpha=1}^N{d\kappa_\alpha\over \kappa_\alpha} 
\exp\left\{i\sum_{ r \in\N}\sum_{ j \in\P}c_{ j  r }W_ j \cdot Z_ r  \right\}
f^{A_1A_2\ldots A_N}\left[\prod_{\alpha=1}^N {d\rho_\alpha\over \rho_\alpha-\rho_{\alpha+1}}\right]\Big/ dg,\cr
\end{align}
where
\be c_{js} ={\kappa_j\over \kappa_s}\prod_{{ r \ne s; r \in\N}}{\rho_j-\rho_ r \over \rho_s-\rho_ r },\label{cjs2}\ee
which we shall relate to (\ref{cjs}) in section 3.

The action of the group GL(2,$\Rop$) is defined by
\be
\rho_\alpha\mapsto {a\rho_\alpha+b\over c\rho_\alpha+d}, \qquad \kappa_\alpha\mapsto (c\rho_\alpha+d)^{n-1} \kappa_\alpha,
\label{GL2}\ee
which leaves $c_{js}$ invariant.

We transform to momentum space by applying 
$$\int \exp\left\{-i\sum_{ j \in\P}\bar\omega_ j \cdot\pi_ j +i\sum_{ r \in\N}\bar\pi_ r \cdot\omega_ r \right\}
\prod_{ j \in\P}d^2\bar\omega_ j \prod_{ r \in\N}d^2\omega_ r, $$
giving
\begin{align}
M^{\epsilon_1\ldots\epsilon_N} = f^{A_1A_2\ldots A_N}
\int \prod_{j\in\P}\delta^2&\left(\pi_j - \sum_{r\in\N}c_{jr}\pi_r\right)
\prod_{ s \in\N}\delta^2\left(\bar\pi_s +\sum_{i\in\P}\bar\pi_i  c_{is}\right)
\cr &\hskip20pt \times
\left[\prod_{\alpha=1}^N{d\kappa_\alpha d\rho_\alpha\over \kappa_\alpha
(\rho_\alpha-\rho_{\alpha+1})}\right]\Big/ dg.
\end{align}
As we saw in section 1, the delta function conditions imply energy-momentum 
conservation. Fixing a choice of $R,S\in\N$,
\begin{align}
\prod_{ j \in\P}&\delta^2\left(\pi_ j -\sum_{ r \in\N}c_{ j  r }\pi_ r \right)
\prod_{s\in\N}\delta^2\left(\bar\pi_s +\sum_{i\in\P}\bar\pi_i  c_{is}
\right)\cr
&=\langle R,S\rangle^2\delta^4\left(\sum_{\alpha=1}^N\bar\pi_\alpha\pi_\alpha\right)\prod_{ j \in\P}\delta^2\left(\pi_ j -\sum_{ r \in\N}c_{ j  r }\pi_ r \right)
\prod_{ s \in\N'}\delta^2\left(\bar\pi_ s +\sum_{ i \in\P}\bar\pi_ i  c_{ i  s }\right),\label{pa}\cr
\end{align}

so that we are left with $2N-4$ relations from the remaining delta functions 
to determine the $2N-4$ effective variables amongst the $\kappa_\alpha,\rho_\alpha$, 
after allowing for M\"obius and scale invariance. The dependence of the 
$mn=n(N-n)$ variables $c_{ir}$ on the $2N-4$ effective variables 
 $\kappa_\alpha,\rho_\alpha$ implies the existence of 
$$N_R=mn-(2N-4)=(m-2)(n-2)$$
(nonlinear) relations between the $c_{ir}$, say $\C_K(c)=0, 1\leq K\leq N_R$. 
{\it E.g.} for $N=6, m=n=3, \P=\{i,j,k\}, \N=\{r,s,t\},$ there is one relation
 from (\ref{C'}):
\begin{align}
c_{ir}c_{js}c_{kt}c_{jr}c_{ks}&c_{it}+c_{jr}c_{ks}c_{it}c_{kr}c_{is}c_{jt}+c_{kr}c_{is}c_{jt}c_{ir}c_{js}c_{kt}\cr
&=c_{kr}c_{js}c_{it}c_{jr}c_{is}c_{kt}+c_{jr}c_{is}c_{kt}c_{ir}c_{ks}c_{jt}+c_{ir}c_{ks}c_{jt}c_{kr}c_{js}c_{it}.\label{constraint}\cr
\end{align}
Now,
$$dg={d\rho_Rd\rho_S\over (\rho_R-\rho_S)^2}{d\kappa_Rd\kappa_S\over \kappa_R\kappa_S},$$
so that 
\be
M^{\epsilon_1\ldots\epsilon_N}
= f^{A_1A_2\ldots A_N}\delta^4\left(\sum_{\alpha=1}^N\bar\pi_\alpha\pi_\alpha\right)\M^{\epsilon_1\ldots\epsilon_N},\ee
with
\begin{align}
\M^{\epsilon_1\ldots\epsilon_N}
= \langle R,S\rangle^2(\rho_R-\rho_S)^2\int 
\prod_{ j \in\P}&\delta^2\left(\pi_ j -\sum_{ r \in\N}c_{ j  r }\pi_ r \right)
\prod_{s\in\N'}\delta^2\left(\bar\pi_s +\sum_{i\in\P}\bar\pi_i c_{is }\right)
\cr
&\times
\prod_{\alpha=1}^N{1\over ( \rho_\alpha-\rho_{\alpha+1})}\prod_{\alpha=1\atop 
\alpha\ne R,S}^N{d\kappa_\alpha d\rho_\alpha\over \kappa_\alpha}.\label{dagg}\cr
\end{align}
Note that if we chose $I,J\in\P$ rather than $R,S\in\N$,
\begin{align}
\prod_{ j \in\P}\delta^2&\left(\pi_ j -\sum_{ r \in\N}c_{ j  r }\pi_ r \right)
\prod_{ s \in\N}\delta^2\left(\bar\pi_ s +\sum_{ i \in\P}\bar\pi_ i  c_{ i  s }\right)\cr
=&[I,J]^2\delta^4\left(\sum_{\alpha=1}^N\bar\pi_\alpha\pi_\alpha\right)\prod_{ j \in\P'}\delta^2\left(\pi_ j -\sum_{ r \in\N} c_{ j  r }\pi_ r \right)
\prod_{ s \in\N}\delta^2\left(\bar\pi_ s +\sum_{ i \in\P}\bar\pi_i c_{is }
\right),\cr
\end{align}
and we have similar formulae to the above but with $[I,J]^2$ replacing $\langle R,S\rangle^2$.

\section{\bf The Amplitude as an Integral over Constraints}

In order take into account the constraints $\C_K$
in evaluating the amplitude (\ref{dagg}), we will rewrite  
the world sheet integration on $\rho$ and $k$ as integrals over a set of 
independent link variables. 

In (\ref{dagg}), we have
\begin{align}
c_{js} &={\kappa_j\over \kappa_s}\prod_{{ r \ne s\atop  r \in\N}}
{\rho_j-\rho_ r \over \rho_s-\rho_ r },\qquad j\in \P,\; s\in\N,\cr
\end{align}
but it is convenient to change variables, defining
$$k_j=\prod_{ r \in\N}(\rho_j-\rho_ r )\kappa_j,
\qquad k_s=\prod_{{ r \ne s\atop  r \in\N}}(\rho_s-\rho_ r )\kappa_s, \qquad j\in\P,\;s\in\N, $$
so that 
\be
c_{js}  ={k_j\over k_s} {1\over\rho_j-\rho_s}.\label{c}
\ee
Then (\ref{dagg}) is left unchanged if we replace $\kappa_\alpha$ by 
$k_\alpha$, and the action of the invariance group $GL(2,\Rop)$ is now given by
$$
\rho_\alpha\mapsto {a\rho_\alpha+b\over c\rho_\alpha+d},\quad
 k_j\mapsto (ad-bc)k_j/(c\rho_j+d),\quad  k_r\mapsto k_r(c\rho_r+d), \qquad \alpha\in\A,\;j\in \P,\; s\in\N,
$$
$\A=\P\cup\N$.
Writing
\be
f_{ir}(c)=\langle i,r\rangle-\sum_{s\in\N}c_{is}\langle s,r\rangle,\qquad 
f_{rt}(c)=[r,t]+\sum_{i\in\P}[r,i]c_{it},\qquad r=R,S,\; i\in\P,\; t\in\N',
\label{f}
\ee
in the expression (\ref{dagg}) we have 
\begin{align}
\langle R,S\rangle^2\prod_{i\in\P}\delta^2&\left(\pi_i-\sum_{ r \in\N}c_{ir}
\pi_r\right)
\prod_{t\in\N'}\delta^2\left(\bar\pi_t+\sum_{i\in\P}\bar\pi_i  c_{it}\right)
\cr
&=K_1\prod_{i\in\P\atop r= R,S}\delta^2\left(f_{ir}(c)\right)
\prod_{t\in\N'\atop r= R,S}\delta^2\left(f_{rt}(c)\right)\equiv K_1
\delta^{N'}\left(f(c)\right),
\end{align}
where $K_1=\langle R,S\rangle^{m+2}[R,S]^{n-2}$.

Use $\varrho_\ell, 1\leq\ell\leq N',$ to denote generically $\{\rho_\alpha,k_\alpha:1\leq \alpha\leq N, \alpha\ne R,S\}$ and divide the $mn$ variables $c_{ir}$ into two subsets:
$c'_\ell, 1\leq \ell\leq N',$ and $c''_K, 1\leq K\leq N_R,$ ({\it e.g.}, fix $I,J\in\P,$ in addition to $R,S\in\N$, and take the $c'$ subset to 
consist of  $\{c_{iR},c_{iS},c_{Ir},c_{Jr}:i\in\P,r\in\N'\}$, 
and the $c''$ subset to consist of $\{c_{ir}:i\ne I,J,\; r\ne R,S\})$. 
Then (\ref{dagg}) can be written \be
\M^{\epsilon_1\ldots\epsilon_N}=K_1\int \Psi(\varrho)\; \delta^{N'}( f)d^{N'}
\varrho\label{M}
\ee
where $f\equiv \tilde f(\varrho) = f(c(\varrho))$
and
\be\Psi(\varrho)=(\rho_R-\rho_S)^2\prod_\alpha{1\over \rho_\alpha-
\rho_{\alpha+1}}\prod_{\alpha\ne R,S}{1\over k_\alpha}.\label{PSI}\ee
In principle, we could use the $N'$ delta functions to perform the 
$N'$ integrals over $\varrho_\ell$ 
and then calculate the amplitude as a function of the momenta by solving the equations $
(\ref{f})$ to give $\rho$ and $k$ in terms of the momenta and then 
substitute for them in 
\be\M^{\epsilon_1\ldots\epsilon_N}=K_1 \Psi(\varrho)\left|{\partial(f)\over 
\partial (\varrho)}\right|^{-1},\label{nf}\ee
but this is not a calculationally convenient way forward. 

Instead, we seek to rewrite (\ref{M})
first as an integral over all the $mn$ variables $c_{ir}$ and then to use the  delta functions $\delta^{N'}\left(f(c)\right)$
to perform $N'$ of these integrations to leave an integral over $N_R$ 
variables corresponding to the constraints $\C_K$. We can use these $N_R$ 
constraints to express the $N_R$ variables $c''_K$ as functions of the 
remaining $N'$ variables $c'_\ell$, $c''=\hat c''(c')$ and thus obtain 
$N'$ functions $\hat f(c')=f(c',\hat c''(c'))$; these are the functions we 
would obtain if we used the $N'$ equations $(\ref{c})$, corresponding to the 
$c'$ to express $\rho, k$ as functions of the $c'$.
\begin{align}
\int \Psi(\varrho)\, \delta^{(N')}(\tilde f)\, d^{N'}\varrho&=\int \Psi(\varrho)
\delta^{N'}(\hat f)\left|{\partial (c')\over \partial (\varrho)}\right|^{-1}d^{N'}c'\cr
&=\int \Psi(\varrho)\left|{\partial (c')\over \partial (\varrho)}\right|^{-1}\delta^{N'}(\hat f)\delta^{N_R}(\C)\left|{\partial (\C)\over \partial (c'')}\right|d^{mn}c\cr
&=\int \Psi(\varrho)\left|{\partial (c')\over \partial (\varrho)}\right|^{-1}\left|{\partial (\C)\over \partial (c'')}\right|\delta^{N'}( f)\delta^{N_R}(\C)d^{mn}c\cr
&=\int \Psi(\varrho)\left|{\partial (c')\over \partial (\varrho)}\right|^{-1}\left|{\partial (\C)\over \partial (c'')}\right|\delta^{N'}( f)\delta^{N_R}(\hat\C)d^{mn}c\cr
&=\int \Psi(\varrho)\left|{\partial (c')\over \partial (\varrho)}\right|^{-1}\left|{\partial (\C)\over \partial (c'')}\right|
\left|{\partial (f)\over \partial (c')}\right|^{-1}\delta^{N_R}(\hat\C)
d^{N_R} c''\cr
\end{align}
with $\hat\C(c'')= \C(\hat c'(c''),c'')$, where the functions 
$c'=\hat c'(c'')$  are obtained by using the $N'$ equations 
$f(c',c'')=0$ to express $c'$ in terms of $c''$. 
The Jacobian of $f$ with respect to $c'$ is a constant, dependent only on momenta rather than the $c_{ir}$, because the $f$ are linear; the value of the constant depends
on the choice of the $c'_\ell$. If we use the choice $\{c_{iR},c_{iS},c_{Ir},
c_{Jr}:i\in\P,r\in\N'\}$ for $c'$, 
$$\left|{\partial (f)\over \partial (c')}\right|= \langle R,S\rangle^{2m} 
[R,S]^{n-2}[I,J]^{n-2}.$$
Then
\be
\M^{\epsilon_1\ldots\epsilon_N}=K\int  F(c'')\delta^{N_R}(\hat\C)d^{N_R} c''\label{M'}
\ee
with
\be
F(c'')=\Psi(\varrho)\left|{\partial (c')\over \partial (\varrho)}\right|^{-1}
\left|{\partial (\C)\over \partial (c'')}\right|\label{F}
\ee
and
$$K=K_1\left|{\partial (f)\over \partial (c')}\right|^{-1}=\langle R,
S\rangle^{2-m}[I,J]^{2-n}.$$

We can integrate (\ref{M'}) to obtain 
\be
\M^{\epsilon_1\ldots\epsilon_N}=K\sum   F(c'')\left|{\partial (\hat\C)\over 
\partial (c'')}\right|^{-1},\label{M''}
\ee
where the sum is over at least some of the simultaneous solutions of the $N_R$ 
constraint equations $\hat C_K(c'')\equiv C_K(\hat c'(c''),c'')=0$. 
[Note that in (\ref{M''}) the Jacobian is calculated for $\hat\C$, that is 
for $\C$ regarded as a function of $c''$ with $c'$ put equal to 
$\hat c'(c'')$, whereas in (\ref{F}) is for $\C$ with respect to $c''$, with all the $c$ regarded as independent.]
To find a rational answer for the amplitude, in line with the known 
results, we need to sum over all the solutions $c''$ 
with appropriate signs or phases that 
enable the contributions to be combined into a contour integral of the form
\be\M^{\epsilon_1\ldots\epsilon_N}=K\oint  F(c'')\prod_{K} 
{dc''_K\over \hat\C_K}.
\label{M0}\ee
[Here
we understand the notation for the contour integral to include 
appropriate factors of $2\pi i$.] This will become apparent when we discuss 
the 6-point function, with $m=n=3$, \break in detail in section 6. 
But first we will discuss the form of the integrand $F(c)$ in section 4, and 
the parameterization of the general solution for $c_{ir}$ of the 
linear equations $f_\ell=0$ in section 5.  

\section{\bf The Form of the Integrand, $F(c)$}

We can now give a general prescription for the 
integrand $F(c)$ in terms of the link variables, 
working from the string function (\ref{PSI}).
As before, we fix $I,J\in\P$ and $R,S\in\N$, set $\P'=\{k\in\P:k\ne i,j\}$ 
and $\N'=\{t\in\N:t\ne r,s\}$, and chose for the $N_R$ variables $c''$ 
the collection $\{c_{kt}:k\in\P', t\in\N'\}.$  Then
the remaining variables  $c'$ are
$\{c_{iR}, c_{iS}, c_{It}, c_{Jt}: i\in\P, t\in\N'\}$. 
Correspondingly, we take the $N_R$ constraint functions
$\C_K$ to be
\be\C_{kt}\equiv \C^{IJk}_{RSt}, \quad k\in\P', \;t\in\N'.\ee
If $k,k'\in\P'$, and $t,t'\in\N'$,
\be{\partial \C_{kt}\over\partial c_{k't'}}=0\qquad\hbox{unless }k=k'\hbox
{ and } t=t',\ee
so that
\begin{align}
\left| {\partial (\C)\over\partial (c'')}\right|&=\prod_{k\in\P'\atop t\in\N'} {\partial \C_{kt}\over\partial c_{kt}}.\cr
&=\left[c_{IR}c_{JS}-c_{IS}c_{JR}\right]^{N_R}
\prod_{ t\in\N'}\left[c_{It}c_{Jt}\right]^{m-2}
\prod_{k\in\P'}\left[c_{kR}c_{kS}\right]^{n-2}\prod_{k\in\P'\atop t\in\N'}
{1\over c_{kt}},\cr
\end{align}
using the expression we find from (\ref{C'}),
\begin{align}
{\partial \C_{kt}\over\partial c_{kt}}&=c_{kS}c_{IR}c_{JR}(c_{It}c_{JS}-c_{IS}
c_{Jt})-c_{kR}c_{IS}c_{JS}(c_{It}c_{JR}-c_{IR}c_{Jt})\cr
&=c_{It}c_{Jt}c_{kR}c_{kS}(c_{IR}c_{JS}-c_{IS}c_{JR})/c_{kt}.
\end{align}
Also it is not difficult to see that the Jacobian of $c'$ with respect to $\rho,k$ can be written as a product of factors:
\begin{align}
\left|{\partial (c')\over\partial(\rho,k)}\right|&=\left|{\partial (c_{IR},c_{IS})\over\partial(\rho_I,k_I)}\right|
\times\left|{\partial (c_{JR},c_{JS})\over\partial(\rho_J,k_J)}\right| \times\prod_{k\in\P'}
\left|{\partial (c_{k\vphantom{J}R},c_{kS})\over\partial(\rho_{\vphantom{J}k},k_{\vphantom{J}k})}\right|
\times \prod_{t\in\N'}\left|{\partial (c_{It},c_{Jt})\over \partial ( \rho_t,k_{\vphantom{J}t})}\right|\cr
&={k_I^{n-1}k_J^{n-1}\over k_R^mk_S^m} (\rho_R-\rho_S)^m(\rho_I-\rho_J)^{n-2}\prod_{k\in\P'}
{k_k}\times \prod_{t\in\N'}{1\over k_t^3}\cr
&\hskip30pt\times\prod_{\ l\in\P}{1\over (\rho_l-\rho_R)^2(\rho_l-\rho_S)^2}
\prod_{ t\in\N'}{1\over (\rho_I-\rho_t)^2(\rho_J-\rho_t)^2},
\end{align}
using
$$
{\partial (c_{IR},c_{IS})\over\partial(\rho_I,k_I)}=-{k_I\over k_Rk_S}{(\rho_R-\rho_S)\over (\rho_I-\rho_R)^2(\rho_I-\rho_S)^2},$$
$$
{\partial (c_{It},c_{Jt})\over \partial ( \rho_t,k_t)}={k_Ik_J\over (k_t)^3}{(\rho_I-\rho_J)\over (\rho_I-\rho_t)^2(\rho_J-\rho_t)^2}.
$$
So
\begin{align}&
(\rho_R-\rho_S)^2\left|{\partial (c')\over\partial(\rho,k)}\right|^{-1}\prod_{l\in\P}{1\over k_l}\prod_{k\in\N'}{1\over k_k}\cr
&={k_R^{2-m}k_S^{2-m}\over k_I^{2-n}k_J^{2-n}} (\rho_R-\rho_S)^{2-m}(\rho_I-\rho_J)^{2-n}c_{IR}^{-2}c_{IS}^{-2}c_{JR}^{-2}c_{JS}^{-2}\prod_{k\in\P'}c_{kR}^{-2}c_{kS}^{-2}\prod_{ t\in\N'}c_{It}^{-2}c_{Jt}^{-2}\prod_{l\in\P}{ k_l^2}\prod_{u\in\N}{1\over k_u^2},\cr
\end{align}
and, using
\be
{k_rk_s\over k_ik_j}(\rho_i-\rho_j) (\rho_s-\rho_r)
={c_{ir}c_{js}-c_{is}c_{jr}
\over c_{ir}c_{is}c_{jr}c_{js}},\qquad\hbox{for}\; i,j\in\P,\;r,s\in\N,\label{R}
\ee
we have
\begin{align}
&(\rho_R-\rho_S)^2\left|{\partial (c')\over\partial(\rho,k)}\right|^{-1}\left| {\partial (\C)\over\partial (c'')}\right|\prod_{i\in\P}{1\over k_i}\prod_{k\in\N'}{1\over k_k}\cr
={k_I^{n}k_J^{n}\over k_R^{m}k_S^{m}} (\rho_R-\rho_S)^{-m}
&(\rho_I-\rho_J)^{-n}{\left[c_{IR}c_{JS}-c_{IS}c_{JR}\right]^{N_R+2}
\over c_{IR}^{3}c_{IS}^{3}c_{JR}^{3}c_{JS}^{3}}\cr
&\times \prod_{ t\in\N'}c_{It}^{m-3}c_{Jt}^{m-3}
\prod_{k\in\P'}c_{kR}^{n-3}c_{kS}^{n-3}\prod_{k\in\P\atop t\in\N}{1\over c_{kt}}\prod_{l\in\P}{ k_l^2}\prod_{u\in\N}{1\over k_u^2},\cr
\end{align}and finally
\begin{align}
F(c)&=\Psi(\rho,k)\left|{\partial (c')\over \partial (\rho,k)}\right|^{-1}
\left|{\partial (\C)\over \partial (c'')}\right|\cr
&={k_I^{n}k_J^{n}\over k_R^{m} k_S^{m}} (\rho_R-\rho_S)^{-m}
(\rho_I-\rho_J)^{-n}{\left[c_{IR}c_{JS}-c_{IS}c_{JR}\right]^{N_R+2}
\over c_{IR}^{3}c_{IS}^{3}c_{JR}^{3}c_{JS}^{3}}
\prod_{ t\in\N'}c_{It}^{m-3}c_{Jt}^{m-3}
\prod_{k\in\P'}c_{kR}^{n-3}c_{kS}^{n-3}
\cr& \hskip80pt \times\prod_{k\in\P\atop t\in\N}{1\over c_{kt}}
\prod_{l\in\P}{k_l^2}\prod_{u\in\N}{1\over k_u^2}\prod_{\alpha=1}^N
{1\over \rho_\alpha-\rho_{\alpha+1}}.
\label{bigF}\end{align}
We can write this final product in terms of the $c_{ir}$ by using 
(\ref{c}) whenever $\epsilon_\alpha=-\epsilon_{\alpha+1}$; and using (\ref{R}), 
with a factor of $(\rho_R-\rho_S)$ supplied when 
$\epsilon_\alpha=\epsilon_{\alpha+1}=+$, and a factor of $(\rho_I-\rho_J)$ 
supplied when $\epsilon_\alpha=\epsilon_{\alpha+1}=-.$ This will leave 
$(\rho_I-\rho_J)^{-p}(\rho_R-\rho_S)^{-p}$, where $p$ is the number of sign 
changes going from $\epsilon_1$ to $\epsilon_N$ and back to $\epsilon_1$. 
These factors can again be converted using (\ref{R}), yielding a rational 
expression for $F(c)$ of order $6(m-2)(n-2)-mn$. We now give some 
examples and a general prescription.
\vs

(a) For $m=n$ and $(\epsilon_1,\epsilon_2,\epsilon_3,\epsilon_4,\ldots
\epsilon_{2n-1},\epsilon_{2n})=(+,-,+,-,\ldots,+,-)$,
\begin{align}F(c)&=\left[c_{IR}c_{JS}-c_{IS}c_{JR}\right]^{(n-2)(n-3)}
\prod_{ t\in\N'}c_{It}^{n-3}c_{Jt}^{n-3}
\prod_{k\in\P}c_{kR}^{n-3}c_{kS}^{n-3}\prod_{k\in\P\atop t\in\N}'
{1\over c_{kt}},
\end{align}
where the prime on the last product indicates that terms $1/c_{kt}$ should be omitted when $k,t$ are adjacent.
\vs 

(b) For $(\epsilon_1,\ldots,\epsilon_m,\epsilon_{m+1},\ldots\epsilon_{m+n})
=(+,\ldots,+,-,\ldots,-)$, where we choose the labeling\break
$\epsilon_{i}=+$, $1\leq i\leq m$; 
$\epsilon_{ r }=-$, $m+1\leq  r \leq N$; and  $I=1,J=m,R=m+1,S=N$, 
we have

\begin{align}
F(c)=- c_{IS}c_{JR}&\left[c_{IR}c_{JS}-c_{IS}c_{JR}\right]^{N_R+1}
\prod_{i=1}^{m-1}{1\over\left[c_{iR}c_{i+1,S}-c_{i+1,R}c_{iS}\right]}
\cr &\times \prod_{ r =m+1}^{N-1}{1\over \left[c_{I r }c_{J, r+1}
-c_{I, r+1}c_{J r }\right]}\,
\prod_{ t\in\N'}c_{It}^{m-1}c_{Jt}^{m-1}
\prod_{k\in\P'}c_{kR}^{n-1}c_{kS}^{n-1}\prod_{k\in\P\atop t\in\N}{1\over c_{kt}}
.\cr
\end{align}

(c) In general, if $(\epsilon_1,\ldots,\epsilon_N)$, begins with $\epsilon_1=+$ and ends with $\epsilon_N=-$ and comprises $p$ strings with $\epsilon_\alpha=+$ and, therefore, $p$ strings with 
 $\epsilon_\alpha=-$, then, up to sign, $F(c)$ is given by
\begin{align}
F(c)
=\left[c_{IR}c_{JS}-c_{IS}c_{JR}\right]^{N_R-p+2}&c_{IR}^{p-3}c_{IS}^{p-3}
c_{JR}^{p-3}c_{JS}^{p-3}
\prod_{ t\in\N'}c_{It}^{m-3}c_{Jt}^{m-3}\cr
&\hskip-20pt \times
\prod_{k\in\P'}c_{kR}^{n-3}c_{kS}^{n-3}\prod_{k\in\P\atop t\in\N}
{1\over c_{kt}}\prod_{\alpha=1}^N d_{\alpha ,\alpha+1},
\end{align}
where
$$d_{ir}=c_{ir},\quad d_{ri}=c_{ir},\quad d_{ij}={c_{iR}c_{jS}c_{jR}c_{iS}
\over c_{iR}c_{jS}-c_{jR}c_{iS}},\quad d_{rs}={
c_{Ir}c_{Js}c_{Is}c_{Jr}\over c_{Ir}c_{Js}-c_{Is}c_{Jr}},\qquad i,j\in\P,\; 
r,s\in\N.$$

Note, one may obtain different expressions for the  integrands  using the identity
$${c_{iR}c_{jS}-c_{jR}c_{iS}\over c_{IR}c_{JS}-c_{IS}c_{JR}}=    {c_{ir}c_{js}-c_{is}c_{jr}\over c_{Ir}c_{Js}-c_{Is}c_{Jr}}\times
{c_{iR}c_{jS}c_{jR}c_{iS}c_{Ir}c_{Js}c_{Is}c_{Jr}\over c_{ir}c_{js}c_{is}c_{jr}c_{IR}c_{JS}c_{IS}c_{JR}}.$$
 For given $m$ and $n$, the expressions for $F(c)$ for different orderings of the helicities are related by the transformations given in Appendix  \ref{sect:Intpn}.

\vss
\section{\bf Parameterization of the General Solution of Linear 
Constraints on $c_{ir}$}

In this section we will parameterize the general solution to the linear 
constraints (\ref{L1}), (\ref{L2}) on the link variables in order to 
express them in terms of suitable independent variables over which to 
perform the multi-dimensional contour integral to obtain the amplitudes.
As remarked in section 1, the $2N$ linear equations (\ref{L1}), (\ref{L2}) 
imply energy-momentum conservation. Thus they typically provide $N'=2N-4$ 
constraints on the variables $c_{ir}$, leaving 
$N_R=mn-N'$ degrees of freedom. These remaining degrees of freedom are 
determined by the $N_R$ independent constraints $\C_K$ that follow from the 
requirement that $c_{ir}$ be of the form (\ref{c}), and we shall discuss the 
form of these non-linear constraints in this section. 

A solution to the linear equations (\ref{L1}), (\ref{L2}) is always provided by 
$c_{ir}=a_{ir}$, where
\be
a_{ir}={1\over p^2}\sum_{{j}\in\P}\langle i, {j}\rangle [{j},r]=-{1\over p^2}
\sum_{{s}\in\N}\langle i, {s}\rangle [{s},r],
\label{a}\ee
using energy-momentum conservation, and
\be
p=\sum_{{j}\in\P}p_{j}=-\sum_{r\in\N}p_r,\qquad
p^2=\sum_{i<{j}\atop {i}, {j}\in\P}\langle {i},{j}\rangle[{i},{j}]=\sum_{r<{s}\atop r, {s}\in\N}\langle r,{s}\rangle[r,{s}].\label{T1}\ee

To show that $c_{ir}=a_{ir}$ satisfies (\ref{L1},\ref{L2}), 
first note that 
$$
\langle i,j\rangle \pi_k +\langle j,k\rangle \pi_i +\langle k,i\rangle \pi_j =0,\qquad \hbox{for any } i,j,k,
$$
because, taking the angle bracket with any vector $\pi_l$,
$$
\langle i,j\rangle \langle k,l\rangle +\langle j,k\rangle \langle i,l\rangle=\langle k,l\rangle\langle i,j\rangle  -\langle k,j\rangle \langle i,l\rangle
=-\langle k,i\rangle\langle j, l \rangle.
$$
Similarly 
\be
[ r,s] \bpi_t +[ s,t] \bpi_r +[ t,r] \bpi_s =0,\qquad \hbox{for any } r,s,t.
\label{T2}\ee
So
\begin{align}
\sum_{r\in\N}a_{ir}\pi_r
&=-{1\over p^2}\sum_{r,{s}\in\N}\langle i, {s}\rangle [{s},r]\pi_r\cr
&=-{1\over 2p^2}\sum_{r,{s}\in\N}[{s},r]\left(\langle i, {s}\rangle \pi_r
-\langle i, r\rangle\pi_{s}\right)\cr
&={1\over 2p^2}\sum_{r,{s}\in\N}[{s},r]\langle {s}, r\rangle \pi_i
= \pi_i,
\end{align}
establishing (\ref{L1}); (\ref{L2}) follows similarly.

For convenience write
\be A_{ir}=p^2a_{ir}.\label{bigA}\ee
Then, for $i,j\in\P$, $r,s\in\N$,

\begin{align}
A_{ir}A_{js}-A_{is}A_{jr}&=\sum_{{u},{v}\in\N}\left(\langle i, {u}\rangle [{u},r]\langle j, {v}\rangle [{v},s]
-\langle i, {u}\rangle [{u},s]\langle j, {v}\rangle [{v},r]\right)\cr
&=\sum_{{u},{v}\in\N}\langle i, {u}\rangle\langle j, {v}\rangle\left( [{u},r] [{v},s]
- [{u},s] [{v},r]\right)\cr
&=[r,s]\sum_{{u},{v}\in\N}\langle i, {u}\rangle\langle j, {v}\rangle[{u},{v}] =[r,s]\sum_{{u}<{v}\atop {u},{v}\in\N}\left(\langle i, {u}\rangle\langle j, {v}\rangle-\langle i, {v}\rangle\langle j, {u}\rangle\right)[{u},{v}] \cr
&=p^2[r,s]\langle i, j\rangle\label{AA}
\end{align}
and so, for $i,j,k\in\P$, $r,s,t\in\N$,
\begin{align}
(p^2)^3\left|\begin{matrix} a_{ir}&a_{is}&a_{it}\cr a_{jr}&a_{js}&a_{jt}\cr 
a_{kr}&a_{ks}&a_{kt}\cr\end{matrix}\right|
&=(A_{ir}A_{js}-A_{jr}A_{is})A_{kt}+(A_{jr}A_{ks}-A_{kr}A_{js}) A_{it}+(A_{kr}A_{is}  -A_{ir}A_{ks})A_{jt}\cr
&=p^2{[ r,s]} \left(\langle i,j\rangle A_{kt}+\langle j,k\rangle A_{it}
+\langle k,i\rangle A_{jt}\right)=0,
\end{align} using (\ref{T1}). 

Since this determinant vanishes for any  $i,j,k\in\P$, $r,s,t\in\N$, this 
implies that the matrix 
\begin{align}
\left(\begin{matrix} a_{i_1r_1}&a_{i_1r_2}&\ldots &a_{i_1r_n}\cr
a_{i_2r_1}&a_{i_2r_2}&\ldots &a_{i_2r_n}\cr
\vdots&\vdots&&\vdots\cr
a_{i_mr_1}&a_{i_mr_2}&\ldots &a_{i_mr_n}\cr\end{matrix}\right)
\label{A}\end{align}
has rank 2. In fact, this is evident from the fact that $a_{ir}$ is defined 
 in (\ref{a}) as the product of 
 the $m\times m$ dimensional matrix 
$\langle i,j\rangle$ and  the $m\times n$ dimensional matrix
$[j,r]$; because the $\pi_i$ are two-dimensional, the matrix 
$\langle i,j\rangle$ has rank at most two and the rank of $a_{ir}$ 
can not be larger.
For $m,n>2$, this condition is different from the rather more unusual 
condition (\ref{C}).

The condition that the matrix (\ref{C}) have rank 2 is sufficient as well as necessary for  $c_{jr}$ to be of the form
$$c_{jr}={k_j\over k_r(\rho_j-\rho_r)}.$$
We can prove this by induction on the size of (\ref{C}). 
Suppose (\ref{C}) is of size $m\times n$, has rank 2 and that the result holds for matrices of this size; consider adding an additional column $x_i=(c_{i,n+1})^{-1}, 1\leq i\leq m$, while leaving the rank of the matrix at 2. Then

\begin{align}
x_i&=\lambda {k_1\over k_i} (\rho_i-\rho_1)+\mu {k_2\over k_i} 
(\rho_i-\rho_2)
={k_{n+1}\over k_i} (\rho_i-\rho_{n+1}),
\end{align}
where
$$
k_{n+1}=\lambda k_1+\mu k_2,
\qquad\rho_{n+1} = {\lambda k_1\rho_1+\mu k_2\rho_2\over \lambda k_1+\mu k_2},
$$
so that the elements of  the additional column  are 
$$c_{i,n+1}={k_i\over k_{n+1}(\rho_i-\rho_{n+1})},
$$
 which is of the required form.
 
For $m=2$ or $n=2$, (\ref{L1}), (\ref{L2}) determine $c_{ir}$ uniquely, 
and it is straightforward then to check that $c_{ir}=a_{ir}$ provides 
the well known MHV amplitudes. For $n=2$, $\N=\{R,S\}$,
 $$ c_{iR}=-{1\over p^2}\langle i, S\rangle [S,R]={\langle i,S\rangle\over \langle R,S\rangle},\qquad
 c_{iS}={\langle i,R\rangle\over \langle S,R\rangle},\qquad\hbox{as }p^2= \langle R,S\rangle[R,S]$$
in this case. Then 
$$c_{i,R}c_{i+1,S}-c_{i,S}c_{i+1,R}={1\over p^2}\langle i,i+1\rangle [R,S]={\langle i,i+1\rangle\over \langle R,S\rangle}.$$
Then, from the formulae at the end of section 4, 
 $$\M^{+\ldots+--}=\langle R,S\rangle^4 \left/\prod_{\alpha=1}^N\langle \alpha,\alpha+1\rangle\right., $$
since there are no integrations to perform in this case.
Similarly for $m=2$, $\P=\{I,J\}$,
$$c_{Ir}={1\over p^2}\langle I, J\rangle [J,r]=-{[J,r]\over [J,I]},\qquad
c_{Jr}=-{[I,r]\over [I,J]},\qquad\hbox{as }p^2= \langle I,J\rangle[I,J]$$
in this case, leading to the familiar form of the amplitude  for this case. 
 
For $m, n>2$ we must add to $a_{ir}$ a solution of the corresponding 
homogeneous linear equations,
\begin{align}
\sum_{r\in\N} \hat c_{jr}\pi_r&=0,\qquad j\in\P\label{H1}\\
\sum_{{i}\in\P}\bar\pi_{i} \hat c_{{i} s}&=0, \qquad s\in\N,\label{H2}
\end{align}
in order to obtain a solution of (\ref{L1}), (\ref{L2}) 
that has the property that (\ref{C}) is of rank 2, 
when \be c_{ir}=a_{ir}+\hat c_{ir}.\label{fullc}\ee
The $\hat c_{ir}$ lie in an $N_R$-dimensional space. For $m=n=3$, 
$\P=\{i,j,k\}$, $\N=\{r,s,t\}$, this space is one-dimensional and parametrized 
by $$\hat c_{ir}={\beta\over 4p^2} \epsilon_{i{j}{k}}[{j},{k}]
\epsilon_{r{s}{t}}\langle {s},{t}\rangle.$$
The single constraint  from (\ref{C}), $\C_1 \equiv
\C^{ijk}_{rst}=0$ provides 
a quartic equation to determine $\beta$, which we shall discuss further in section 6. 

For general $m, n\geq 3$, the general solution to (\ref{H1}), (\ref{H2}) 
is provided by 
$$\hat c_{ir}={1\over 4} \sum_{j,k\in\P\atop s,t\in \N}\beta^{ijk}_{rst}[j,k]\langle s,t\rangle,$$
where $\beta^{ijk}_{rst}$ is antisymmetric under permutations of $i,j,k$ and 
also under the permutations of $r,s,t$; it follows from (\ref{T1}), (\ref{T2})
that this 
satisfies (\ref{H1}), (\ref{H2}). Because there are only 
$(m-2)(n-2)$ independent solutions to (\ref{H1}), (\ref{H2}) there is some 
arbitrariness in the choice of $\beta^{ijk}_{rst}$ for a given solution $c_{ir}$, e.g. 
$$\beta^{ijk}_{rst}\mapsto\beta^{ijk}_{rst} + [i,l]\gamma^{jk}_{rst} +[j,l]\gamma^{ki}_{rst} +[k,l]\gamma^{ij}_{rst},$$
where $\gamma^{ij}_{rst}$ is antisymmetric in ${i},{j}$ and in $r,{s},{t}$, and $\bpi_l$ is arbitrary, leaves the solution $\hat c_{ir}$ unchanged. 
\vs
We can express each of the $\hat c_{ir}$ as a linear combination of the $N_R$ components 
$$\hat c_{kt}=\beta_{kt}[I,J]\langle R,S\rangle,\qquad k\in\P', t\in\N'$$
by
\begin{align}
\hat c_{It}=\sum_{k\in\P'}\beta_{kt}[J,k]\langle R,S\rangle,\quad\hat c_{kR}=\sum_{t\in\N'}\beta_{kt}[I,J]\langle S,t\rangle,  \quad
\hat c_{IR}=\sum_{k\in\P'\atop t\in\N'}\beta_{kt}[J,k]\langle S,t\rangle,\cr
\hfil k\in\P',\;t\in\N',
\end{align}
and similar expressions for $\hat c_{kS}, \hat c_{Jt}, 
\hat c_{IS}, \hat c_{JR}, \hat c_{JS}.$ This is equivalent to taking 
$\beta^{ijk}_{rst}=0$
unless exactly one of $i,j,k$ is in $\P'$ and exactly one of $r,s,t$ is in $\N'$. In this case, the only possibly nonzero components of $\beta^{ijk}_{rst}$ are
\be\beta^{IJk}_{RSt}=\beta_{kt}, \qquad k\in\P',\; t\in\N'\label{bkt}\ee
and those related to these by antisymmetry under permutations of  $I,J,k$ or of $R,S,t$. 
\be\beta^{ijk}_{rst}=\epsilon_{ijk}\epsilon_{rst}\beta,\qquad\hbox{for}\quad m=n=3,\label{b6}\ee
and
\be\beta^{ijk}_{rst}=\beta^{ijk}\epsilon_{rst},\qquad\hbox{for}\quad m>3, n=3,\label{bnmhv}\ee
and, in this case, we can specify that the only nonzero components of $\beta^{ijk}$ are the $m-2$ components of the form $\beta^{IJk}=\beta_k$, $k\in\P'$ and those obtained by permuting $I,J,k$.

Then, from (\ref{fullc}), we have,
\be c_{ir}=a_{ir}+{1\over 4p^2} \sum_{j,k\in\P\atop s,t\in\N}\beta^{ijk}_{rst}
[j,k]\langle s,t \rangle,\label{cform}\ee
where the parameters 
$\beta^{ijk}_{rst}$ are zero unless they are related to 
$$\beta^{IJk}_{RSt}=\beta_{kt} \qquad k\in\P',\;
t\in\N',$$
by permutation of $i,j,k$ and $r,s,t$. Then, from (\ref{M0}), the amplitude becomes  
\be\M^{\epsilon_1\ldots\epsilon_N}=\tilde K\oint  F(c)\prod_{k\in\P'\atop 
t\in\N'} {d\beta_{kt}\over \C_{kt}(\beta)},\label{contour}\ee
with
\be\tilde K=[ I,J]^{(m-3)(n-2)}\langle R,S
\rangle^{(m-2)(n-3)}\left(p^2\right)^{ (m-2)  (2-n)},\label{defK}\ee
and $\C_{kt}=\C^{IJk}_{RSt}$  provide the $N_R$ constraints $\C_K$.

\section{\bf NMHV 6-point Functions.}
\nobreak
We will exemplify our analysis by computing all NMHV 6-point gluon tree amplitudes. Following \cite{ACCK2}, we then find
equivalent but different expressions \cite{H1,H2} for the amplitudes, 
by choosing other equivalent contours and integrands. 

For $m=n=3$, we take $\P=\{i,j,k\}, \N=\{r,s,t\}$; from (\ref{contour}), 
\be\M^{\epsilon_1\ldots\epsilon_6}={1\over p^2}\oint  F(c) 
{d\beta\over \C(\beta)},\label{sixcontour}\ee
with just one constraint $\C_1\equiv\C_{kt}  \equiv\C(\beta)$,
\begin{align}
\C(\beta)=c_{ir}c_{js}c_{kt}&c_{jr}c_{ks}c_{it}+c_{jr}c_{ks}c_{it}c_{kr}
c_{is}c_{jt}+c_{kr}c_{is}c_{jt}c_{ir}c_{js}c_{kt}\cr
&-c_{kr}c_{js}c_{it}c_{jr}c_{is}c_{kt}-c_{jr}c_{is}c_{kt}c_{ir}c_{ks}c_{jt}
-c_{ir}c_{ks}c_{jt}c_{kr}c_{js}c_{it};
\end{align} the link variables are 
$$c_{ir}=a_{ir}+{\beta\over p^2} [ j , k ]\langle  s , t \rangle,$$
where $(i,j,k)$ and $(r,s,t)$ are cyclic, and 
$$F(c)
={k_i^2k_j^2k_k^2\over k_r^2k_s^2k_t^2}\prod_{k\in\P\atop t\in\N}{1\over c_{kt}}\prod_{\alpha=1}^6      {1\over \rho_\alpha-\rho_{\alpha+1}}.$$

For the three different orderings of the NMHV 6-point functions,
the integrands are given as in section 4.

(a) for the case $(i,t,j,r,k,s)\equiv (+,-,+,-,+,-)$,
$$F^{+-+-+-}(c)=-{1\over c_{ir}c_{js}c_{kt}}; $$

(b) for the case $(k,i,s,t,j,r)\equiv (+,+,-,-,+,-)$
$$F^{++--+-}(c)={c_{is}\over(c_{ks}c_{it}-c_{kt}c_{is})c_{ir}c_{js}};$$

(c) for the case $(i,j,k,r,s,t)\equiv (+,+,+,-,-,-)$
$$F^{+++---}(c)=-{c_{js}\over(c_{ir}c_{js}-c_{is}c_{jr})(c_{js}c_{kt}-c_{jt}c_{ks})}.$$

To compute the amplitudes, we examine the explicit form of the
constraint, which leads to a quartic equation for $\beta$. 
Writing
$$V_i=[j,k],\qquad W_r =\langle s,t\rangle,\qquad \bV_i=\langle j,k\rangle,
\qquad \bW_r=[s,t],$$
and similarly for cyclic rotations of $(i,j, k)$ and of $(r,s,t)$, 
we have 
\be c_{ir}=a_{ir}+{\beta\over  p^2}V_iW_r,\label{cir6}\ee
and, from (\ref{AA}), 
\be A_{ir}A_{js}-A_{is}A_{jr}=p^2\bV_k\bW_t.\label{AA6}\ee
From Appendix \ref{sect:sixptrel}, where we list some algebraic relations useful for 
computing the 6-point functions, 
\be c_{ir}c_{js}-c_{is}c_{jr}= \beta\bc_{tk},\label{cc6}\ee
for
\be \bc_{ri}=\ba_{ri}+{\bbeta\over p^2} \bW_r\bV_i,\qquad\ba_{ri}={1\over p^2}\bA_{ri},\qquad\bbeta=1/\beta,\label{be}\ee
\be \bA_{ri}=\sum_{ s \in\N}\langle r, s \rangle[ s ,i]=-\sum_{ j \in\P}\langle r, j \rangle[ j ,i].\label{ba}\ee

In Appendix \ref{sect:sixptrel}, we see that $\det(c)=\beta$ so that $\bc_{ri}$ is the inverse 
of $c_{ir}$, and hence provides 
the general solution of
$$\pi_r=\sum_{ i \in\P} \bc_{ri}\pi_ i ,\qquad \bpi_i=-\sum_{r\in\N}\bpi_ r \bc_{ r  i},$$
the equations for the amplitude with flipped helicities. 
Again, from Appendix \ref{sect:sixptrel},
$$
\bA_{ri}\bA_{sj}-\bA_{si}\bA_{rj}= p^2W_tV_k,$$
and
$$
\bc_{ri}\bc_{sj}-\bc_{si}\bc_{rj}= \bbeta c_{kt}.
$$

Using these properties of $c_{ir}$, the constraint becomes $\C(\beta)=0$, where
\begin{align}
\C(\beta) &=
c_{is}c_{jr}(c_{kt}c_{ir}-c_{kr}c_{it})(c_{js}c_{kt}-c_{jt}c_{ks})
-c_{ir}c_{js}(c_{ks}c_{it}-c_{kt}c_{is})(c_{jt}c_{kr}-c_{jr}c_{kt})\cr
&=
\beta^2[c_{is}c_{jr}\bc_{ri}\bc_{sj}
-c_{ir}c_{js}\bc_{si}\bc_{rj}]\cr
&=
\beta c_{ir}c_{js}c_{kt}-\beta^3\bc_{ri}\bc_{sj}\bc_{tk},
\end{align}
which is manifestly quartic in $\beta$. Writing
$$\C(\beta)=\beta c_{ir}c_{js}c_{kt}-\beta^3\bc_{ri}\bc_{sj}\bc_{tk}
 \equiv C_{(4)}\beta^4+C_{(3)}\beta^3+C_{(2)}\beta^2+C_{(1)}\beta+C_{(0)},$$ 
the coefficients in the quartic are
\begin{align}
C_{(0)}&=-\bv_i\bv_j\bv_k\bw_r\bw_s\bw_t;\cr
C_{(1)}&= a_{ir}a_{js}a_{kt}-\ba_{ri}\bv_j\bv_k\bw_s\bw_t-\ba_{sj}\bv_i\bv_k\bw_r\bw_t
-\ba_{tk}\bv_i\bv_j\bw_r\bw_s;\cr
 C_{(2)}&=a_{ir}a_{js}v_kw_t +a_{js}a_{kt}v_iw_r+a_{ir}a_{kt}v_jw_s
-\ba_{ri}\ba_{sj}\bv_k\bw_t -\ba_{sj}\ba_{tk}\bv_i\bw_r-\ba_{ri}\ba_{tk}\bv_j\bw_s;\cr
C_{(3)}&=a_{ir}v_jv_kw_sw_t+a_{js}v_iv_kw_rw_t+a_{kt}v_iv_jw_rw_s-\ba_{ri}\ba_{sj}\ba_{tk};\cr
C_{(4)}&=v_iv_jv_kw_rw_sw_t,\cr
\end{align}
 where
$$v_i=V_i/p^2,\qquad w_r=W_r/p^2,\qquad \bv_i=\bV_i/p^2,\qquad \bw_r=\bW_r/p^2.$$

Now consideration of particular cases for the momenta 
demonstrates that the roots of $\C(\beta)$ are irrational functions of the 
momenta and that rational results can only be obtained  by summing 
over each of the roots. So,  from (\ref{sixcontour}),  we take 
\begin{align}
\M^{\epsilon_1\ldots\epsilon_6}
&={1\over p^2}\oint_\O {F^{\epsilon_1\ldots\epsilon_6}(c)\over \C(\beta)}
d\beta,
\end{align}
where
$\O$ is taken to be a contour encircling each of these roots once, but none of 
the poles of $F(c)$. We now discuss the evaluation of the amplitude using 
contour manipulation, in the spirit of  \cite{ACCK1,ACCK2}. 

Because the integrand tends to zero as $\beta^{-3}$ or faster,
as $\beta\rightarrow\infty$, we have 
$$\M^{\epsilon_1\ldots\epsilon_6}
=-{1\over p^2}\oint_{\O'} {F^{\epsilon_1\ldots\epsilon_6}(c)\over \C(\beta)}d\beta,$$
where the contour $\O'$ encircles the poles in $\beta$ of $F^{\epsilon_1\ldots\epsilon_6}(c)$ positively, but excludes the zeros of $\C(\beta)$.

(a) For the case $(i,t,j,r,k,s)\equiv (+,-,+,-,+,-),$
$$\M^{+-+-+-}
={1\over 2\pi i}\oint_{\O_a} {d\beta\over  p^2  c_{ir}c_{js}c_{kt}\C(\beta)},$$
where $\O_a$ encircles $\beta = -A_{ir}/V_iW_r$, $\beta = -A_{js}/V_jW_s$ and $\beta = -A_{kt}/V_kW_t$. The integral is the sum of the contributions of these poles of the integrand, and at each of them $\C(\beta) =\beta c_{ir}c_{js}c_{kt}-\beta^3\bc_{ri}\bc_{sj}\bc_{tk}=-\beta^3\bc_{ri}\bc_{sj}\bc_{tk}$, because $c_{ir}c_{js}c_{kt}=0$. Thus, using formulae listed in Appendix \ref{sect:sixptrel}, 
\begin{align}
\M^{+-+-+-}
&=-{1\over 2\pi i}\oint_{\O_a} {d\beta\over  p^2 c_{ir}c_{js}c_{kt}\beta^3
\bc_{ri}\bc_{sj}\bc_{tk}}\label{a1}\\
& = - {[j,k]^4 \langle s,t\rangle^4
\over \langle t|P_{si}|k]\langle s|P_{ti}|j]\, p^2_{jkr} 
\langle i,s\rangle [r,j] \langle i,t\rangle [r,k]}\quad
+\quad\hbox{2 cyclic terms,}\cr\nonumber
\end{align}
where the  two additional cyclic terms are obtained by simultaneously 
cyclically rotating $(i,j,k)$ and $(r,s,t)$, and
where  we write
\be\langle\alpha_1|P_{\beta_1\ldots\beta_l}|\alpha_2]
=\langle\alpha_1,\beta_1\rangle[ \beta_1,\alpha_2]+\dots+\langle\alpha_1,
\beta_l\rangle[ \beta_l,\alpha_2],\label{bracket}\ee
so that
$$\langle t|P_{si}|k]= 
-\langle t|P_{rj}|k],\;\hbox{\it etc.,}$$
and 
\be p_{\alpha_1\ldots\alpha_l}^2=\left(p_{\alpha_1}+\ldots p_{\alpha_l}\right)^2
.\label{defp}\ee

Having written the amplitude in the form (\ref{a1}), we can replace the contour 
$\O_a$ by a contour $\O_a'$ encircling the poles corresponding to 
$\bc_{ri}=0$, $\bc_{sj}=0$ and $\bc_{tk}=0$, {\it i.e.}
$\beta = -\bV_i\bW_r/\bA_{ir}$, $\beta = -\bV_j\bW_s/\bA_{js}$ and 
$\beta = -\bV_k\bW_t/\bA_{kt}$, respectively:
\begin{align}
\M^{+-+-+-}
&={1\over 2\pi i}\oint_{\O_a'} {d\beta\over  p^2 c_{ir}c_{js}c_{kt}\beta^3
\bc_{ri}\bc_{sj}\bc_{tk}}\label{a2}\\
&=-{ \bar A_{ri}^4
\over \langle k|P_{si}|t] \langle j|P_{ti}|s] p^2_{jkr} [i,s] \langle r,j\rangle [ i,t] \langle r,k\rangle}\quad
+\quad\hbox{2 cyclic terms,}\nonumber
\end{align}
where 
$\bar A_{ri}$ is defined in (\ref{ba}).
(\ref{a2}) provides a second, equivalent form for the alternating tree 
amplitude.

(b) For the case $(k,i,s,t,j,r)\equiv (+,+,-,-,+,-),$
$$\M^{++--+-}
=-{1\over 2\pi i}\oint_{\O_b} {c_{is}d\beta\over p^2  c_{ir}c_{js}\beta\bc_{rj}\C(\beta)},$$
where $\O_b$ encircles $\beta = -A_{ir}/V_iW_r$, $\beta = -A_{js}/V_jW_s$ and $\beta = -\bV_j\bW_r/\bA_{rj}$. The integral is the sum of the contributions of these poles of the integrand, and at each of them $\C(\beta) =\beta^2[c_{is}c_{jr}\bc_{ri}\bc_{sj}-c_{ir}c_{js}\bc_{si}\bc_{rj}]=\beta^2c_{is}c_{jr}\bc_{ri}\bc_{sj}$, because $c_{ir}c_{js}\bc_{si}\bc_{rj}=0$ at the poles. Thus

\begin{align}
\M^{++--+-}
&=-{1\over 2\pi i}\oint_{\O_b} {d\beta\over p^2   c_{ir}c_{js}\beta\bc_{rj}c_{jr}\beta^2\bc_{ri}\bc_{sj}}\label{b1}\\
&=-{ [j,k]^4\langle s,t\rangle^3\over   
\langle t|P_{si}|k]\langle i|P_{rk}|j] [k,r]p^2_{jkr}\langle i,s
\rangle [r,j]}
-{ [k,i]^3\langle t,r\rangle^4\over   
\langle t|P_{rj}|k]\langle r|P_{tj}|s] [i,s]p^2_{kis}\langle j,t\rangle 
\langle j,r\rangle}\cr
&\hskip20truemm-{\bar A_{rj}^4\over   
\langle i|P_{rk}|j] \langle r|P_{tj}|s] 
p_{kir}^2\langle k,r\rangle [j,t][s,t]\langle k,i\rangle}.\nonumber
\end{align}
For an alternative form, 
starting from (\ref{b1}), we can replace the contour $\O_b$ by a contour $\O_b'$ encircling the poles corresponding to $\bc_{ri}=0$, $\bc_{sj}=0$ and $c_{jr}=0$, {\it i.e.}
$\beta = -\bV_i\bW_r/\bA_{ir}$, $\beta = -\bV_j\bW_s/\bA_{js}$ and $\beta = -A_{jr}/V_jW_r$, respectively:
\begin{align}
\M^{++--+-}
&={1\over 2\pi i}\oint_{\O_b'} {d\beta\over p^2   
c_{ir}c_{js}\beta\bc_{rj}c_{jr}\beta^2\bc_{ri}\bc_{sj}}\label{b2}\\
&=-{ \bar A_{ri}^4\over   \langle k|P_{si}|t] 
\langle j|P_{rk}|i] 
\langle k,r\rangle p^2_{jkr}[ i,s] \langle r,j\rangle [s,t]}
-{\bar A_{sj}^4\over\langle k|P_{rj}|t] 
\langle s|P_{tj}|r]\langle i,s
\rangle p^2_{kis}[ j,t] [ j,r]\langle k,i\rangle}\cr
&\hskip20truemm-{ [k,i]^4\langle s,t\rangle^4\over\langle j|P_{rk}|i]
\langle s|P_{tj}|r] 
p_{kir}^2[ k,r] \langle j,t\rangle \langle s,t\rangle [ k,i]}.\label{split}
\end{align}

(c) For the case $(i,j,k,r,s,t)\equiv (+,+,+,-,-,-),$
$$\M^{+++---}
={1\over 2\pi i}\oint_{\O_c} {c_{js}d\beta\over p^2  \beta^2\bc_{tk}\bc_{ri}\C(\beta)},$$
where $\O_c$ encircles $\beta = -\bV_k\bW_t/\bA_{tk}$ and $\beta = -\bV_i\bW_r/\bA_{ri}$. The integral is the sum of the contributions of these poles of the integrand, and at each of them $\C(\beta) =\beta c_{ir}c_{js}c_{kt}-\beta^3\bc_{ri}\bc_{sj}\bc_{tk}=\beta c_{ir}c_{js}c_{kt}$, because $\bc_{ri}\bc_{sj}\bc_{tk}=0$ at the poles. Thus
\begin{align}
\M^{+++---}
&={1\over 2\pi i}\oint_{\O_c} {d\beta\over p^2  \beta^2\bc_{tk}\bc_{ri}
\beta c_{ir}c_{kt}}&\label{c1}\\ 
&=-{\bar A_{ri}^3\over\langle j,k\rangle[s,t]
\langle j|P_{ti}|s] p^2_{jkr}[ i,t] 
\langle r,k\rangle} - {\bar A_{tk}^3  
\over\langle j,i\rangle[s,r]\langle j|P_{rk}|s] 
p^2_{jit}[ k,r] \langle t,i\rangle}.\nonumber
\end{align}

To find the alternative form,
starting from (\ref{c1}), we can replace the contour $\O_c$ by a contour $\O_c'$ encircling the poles corresponding to $c_{ir}=0$, $c_{kt}=0$ and $\beta=0$, {\it i.e.}
$\beta = -A_{ir}/V_iW_r$, $\beta = -A_{kt}/V_kW_t$ as well as $\beta = 0$, respectively:

\begin{align}
\M^{+++---}
&=-{1\over 2\pi i}\oint_{\O_c'} {d\beta\over p^2  \beta^2\bc_{tk}\bc_{ri}
\beta c_{ir}c_{kt}}\label{c2}\\ 
&={ [j,k]^3\langle s,t\rangle^3 \over 
A_{ir} 
\langle s|P_{ti}|j] p^2_{jkr}\langle i,t\rangle[r,k]} +
{ [i,j]^3\langle r,s\rangle^3\over  A_{kt} 
\langle s|P_{kr}|j] \;p^2_{ijt}
\langle r,k\rangle [i,t]}
-{(p^2)^3\over \langle ij\rangle  [r,s]\langle j,k\rangle 
[st]   A_{ir} A_{kt}}, 
\nonumber\end{align}
where the $A_{ir}$ is defined in (\ref{bigA}).
\vss
\section{\bf Multiple Constraints and an NMHV 7-point Function.}
\nobreak
We now consider the general situation in which the number of constraints, $N_R>1$. The general form of the amplitude is 
\be\M^{\epsilon_1\ldots\epsilon_N}=\tilde K\oint  F(c)\prod_{k\in\P'\atop 
t\in\N'} {d\beta_{kt}\over \C_{kt}(\beta)},\label{contour2}\ee
where $\tilde K $ is given by (\ref{defK}). We must specify further the multi-dimensional contour in (\ref{contour2}) and
explain how it may be evaluated. We remember, from the end of section 3, that the motivation for (\ref{contour2}), is obtained by considering the result (\ref{M''}) of evaluating the delta function constraints $ \delta^{N_R}(\hat\C)$ in (\ref{M'}). By replacing real integrals over delta functions with contour integrals  around 
corresponding poles, we effectively replaced an expression involving the inverse of the modulus of the Jacobian  of the constraints $\hat\C$, summed over a discrete set of points corresponding to the simultaneous solution of the constraint conditions, with the same expression with the modulus removed, so that the sum over the simultaneous solutions of the constraints now includes the phases of the inverse Jacobian at these points. We saw in section 6, in the case of the $N_R=1$ 6-point function, that this interpretation was forced on us if we were to be able to reproduce the rational form for the amplitude known from gauge theory.

In order to specify (\ref{contour2}) more precisely, we choose an order for the constraints, $\C_K$, $K=1\ldots N_R,$ and order the parameters, $\beta_K$, correspondingly. Then, as in section 5.1 of \cite{ACCK2}, the contribution of a particular simultaneous solution $\beta=\beta^{0}$ of the constraint equations, $\C_K(\beta)=0$, is 
\be\tilde K\oint_{\O_{\beta^0}}  F(c)\left[\prod_{K=1}^{N_R} {1\over \C_K(\beta)}\right]d\beta
=\left.\tilde KF(c)\left[{\partial(\C)\over\partial(\beta)}\right]^{-1}\right|_{\beta=\beta^0},\label{contour3}\ee
where $d\beta=d\beta_1\wedge\ldots\wedge d\beta_{N_R}$ and the contour $\O^0$ is chosen to be a surface of the form $\{\beta: |\C_K(\beta)|=\epsilon, 1\leq K\leq N_R\}$, with its orientation determined by the order of the $\C_K$ (as in \cite{ACCK2}), enclosing $\beta=\beta^{0}$ but no other zero of the $\C_K$.  The integral (\ref{contour3}) is be the residue of the integrand at $\beta=\beta_0$ and it is antisymmetric under independent permutations of the order of the $\C_K$ or of the $\beta_K$, and so symmetric under simultaneous identical permutations of both.

Now, as in the  $N_R=1$ case, we should sum (\ref{contour3}) over all the simultaneous solutions, $\beta=\beta^0$, of the constraints, $\C_K(\beta)=0$, but exclude the contribution of other poles of the integrand arising from $F(c)$. At this point we should note that, with the particular set of constraints we have chosen in section 4, namely $\{\C_{kt}: k\ne I,J,\; t\ne R,S\}$, there are 
always
four  `trivial'  or `spurious' simultaneous solutions of the constraints, namely $c_{IR}=c_{JR}=0$; $c_{IS}=c_{JS}=0$; $c_{IR}=c_{IS}=0$; and  $c_{JR}=c_{JS}=0$. They are introduced when we move from statements about the matrix (\ref{C}), with entries $c_{ir}^{-1}$ to statements about multinomials (\ref{C'}) in the link variables, $c_{ir}$, by multiplying by products of them. They do not correspond to the matrix (\ref{C}) having rank 2; in fact, they are artifacts of the particular choice of the independent set of constraints. These spurious solutions should be excluded from the sum.

The problem again with trying to evaluate this sum directly is that the solutions to the constraints are not all rational individually, only their sum is. So, we seek to use a multidimensional version of the contour manipulation arguments we used in section 6 to evaluate the amplitude to obtain the familiar rational results. As in \cite{ACCK2}, this is provided by the global residue theorem. Consider an $N$-dimensional integral of the form 
\be \oint {\Phi(\beta)d\beta\over \prod_{\alpha=1}^M h_\alpha(\beta)},\ee
where $N<M$. For distinct indices $\alpha_1,\ldots, \alpha_N$, the residue of the integrand at a common zero $\beta^0$ of $h_{\alpha_1},\ldots,h_{\alpha_N}$, assumed to be a simple zero, is 
\be \R(h_{\alpha_1}, \ldots,h_{\alpha_N})=\Phi(\beta^0)\prod_{\alpha\notin A} {1\over h_\alpha(\beta^0)}\left[{\partial (h_{\alpha_1},\ldots,h_{\alpha_N})\over\partial (\beta_1,\ldots,\beta_N)}\right]^{-1}_{\beta=\beta^0},\label{res}\ee
where $A=\{\alpha_1,\ldots,\alpha_N\}$. It may be that the functions $h_{\alpha_1},\ldots,h_{\alpha_N}$ have more than one common zero but we shall assume the set of such simultaneous zeros is finite and, if there is more than one, we shall understand $\R(h_{\alpha_1}, \ldots,h_{\alpha_N})$ to denote the sum of the residues (\ref{res}) at these simultaneous zeros. Suppose  $\Gamma_\ell, 1\leq\ell\leq N$ are disjoint subsets of $\{1,\ldots,M\}$, whose union is the whole set. Then a version \cite{ACCK2} of the global residue theorem states that 
\be \sum_{\alpha_\ell\in\Gamma_\ell}\R(h_{\alpha_1}, \ldots,h_{\alpha_N})=0.\label{gres}\ee

We demonstrate the effectiveness of this for evaluating amplitudes by considering a 7-point NMHV tree with helicities  $(i,r,k,s,l,t,j)\equiv (+-+-+-+)$, so that  $m=4$, $n=3$,  $N_R=2$. We take the choice $(I,J,R,S)=(i,j,r,s)$. In this case there are two constraints, $\C_{kt}, \C_{lt}$, and, as in (\ref{bnmhv}), we can take the two integration variables to be $\beta_k = \beta^{ijk}$, $\beta_l = \beta^{ijl}$,
where, for $r\in\N$,
\be c_{ir}=a_{ir}+( \beta_k [j,k] + \beta_l [j,l])W_r/p^2,\label{cir1}\ee
\be c_{jr}=a_{jr}+(\beta_k [k,i] + \beta_l [l,i])W_r/p^2,\label{cjr2}\ee
\be c_{kr}=a_{kr}+\beta_k [i,j]W_r/p^2,\quad 
c_{lr}=a_{lr}+\beta_l [i,j]W_r/p^2,\quad r\in\N.\label{cktclr}\ee

From (\ref{bigF}),
\be
F(c) = {c_{ir}c_{jt}\over c_{kt}c_{lr}},\label{Fseven}\ee
and so, from (\ref{contour}), the contour expression for the amplitude is of the form
\begin{align}
\M ={[i,j]\over (p^2)^2} \,\oint_\O {c_{ir}c_{jt}\over c_{kt}c_{lr}
\C_{kt}\C_{lt}} d\beta_k d\beta_l.\label{seven}\end{align}
In (\ref{seven}) the contour $\O$ is chosen so that it includes the residue contributions from each of the simultaneous solutions of the constraints $\C_{kt}=\C_{lt}=0$, but excluding the `spurious' solutions. 
$\C_{kt}=\C_{lt}=0$ if $c_{ir}=c_{jr}=0$ or $c_{is}=c_{js}=0$ or 
$c_{it}=c_{jt}=0$; we do not have to consider possible spurious contributions 
from, e.g. $c_{ir}= c_{is}=0$, because the variables $c_{ir},c_{is},c_{it}$ 
are not independent and no two of them can vanish together for general 
momenta, and similar considerations apply to $c_{jr},c_{js},c_{jt}$. 
The residues at  $c_{ir}=c_{jr}=0$ 
and $c_{it}=c_{jt}=0$ vanish because of the presence of $c_{ir}$ and $c_{jt}$, 
respectively, in the numerator of the integrand, so that we only have to 
exclude  the contribution from $c_{is}=c_{js}=0$ explicitly.


Applying the global residue theorem taking $\Gamma_1$ to correspond to $\{\C_{kt},c_{lr}\}$ and $\Gamma_2$ to correspond to $\{\C_{lt},c_{kt}\}$,
we obtain
\be \R(\C_{kt},\C_{lt})+\R(c_{lr},\C_{lt})+\R(\C_{kt},c_{kt})+\R(c_{lr},c_{kt})=0.\label{grt1}\ee
and so, excluding the `spurious' contribution,
\begin{align}
\M&=\R(\C_{kt},\C_{lt})-\R(c_{is},c_{js})\cr
&=-\R(c_{lr},c_{kt})-\R(\C_{kt},c_{kt})-\R(c_{lr},\C_{lt})-\R(c_{is},c_{js})\label{RT1}
\end{align}
The choice of $\Gamma_1$ and $\Gamma_2$ has been made so that $\R(\C_{kt},c_{kt})$ and $\R(c_{lr},\C_{lt})$ are as easy to evaluate as $\R(c_{lr},c_{kt})$ and $\R(c_{is},c_{js})$ are.

Because 
\begin{align}{\cal C}_{kt} &=
c_{ir}c_{js}c_{kt}c_{jr}c_{ks}c_{it}+c_{jr}c_{ks}c_{it}c_{kr}c_{is}c_{jt}
+c_{kr}c_{is}c_{jt}c_{ir}c_{js}c_{kt}\cr
&\hskip30pt 
-c_{kr}c_{js}c_{it}c_{jr}c_{is}c_{kt}-c_{jr}c_{is}c_{kt}c_{ir}c_{ks}c_{jt}
-c_{ir}c_{ks}c_{jt}c_{kr}c_{js}c_{it}\cr
&=c_{is}c_{kt}c^{jk}_{rs}c^{ij}_{tr}-c_{it}c_{ks}c^{jk}_{tr}c^{ij}_{rs},\label{C2}\\
\nonumber\end{align}
where
\be c^{ij}_{rs}=c_{ir}c_{js}-c_{is}c_{jr},\label{cijrs}\ee
we have that
\be \C_{kt}|_{c_{kt}=0}= c_{kr}c_{ks}c_{it}c_{jt}(c_{is}c_{jr}-c_{ir}c_{js}),\label{Ckt0}\ee
from which it follows that we can write $ \R(\C_{kt},c_{kt})$ as a sum of terms corresponding to the factors of $\C_{kt}$ when $c_{kt}=0$. Because, in this case, for general momenta, 
$c_{kr}, c_{ks}\ne 0$ when $c_{kt}=0$, and because of the presence of $c_{jt}$ in the numerator, we only have to consider the factors $c_{it}$ and $c^{ij}_{rs}$, implying
\be \R(\C_{kt},c_{kt})=\R(c_{it},c_{kt})+\R(c^{ij}_{rs},c_{kt}).\label{RCktckt}\ee
Similarly, from 
\be \C_{lt}|_{c_{lr}=0}= c_{ls}c_{lt}c_{ir}c_{jr}(c_{it}c_{js}-c_{is}c_{jt}), \label{Clt0}\ee
we deduce
\be \R(c_{lr},\C_{lt})=\R(c_{lr},c_{jr})+\R(c_{lr},c^{ij}_{st}).\label{RClrclt}\ee
Combining (\ref{RCktckt}) and (\ref{RClrclt}) with (\ref{RT1}), we have $\M$ expressed as a sum of six terms,
\be
\M=-\R(c_{lr},c_{kt})-\R(c_{it},c_{kt})-\R(c^{ij}_{rs},c_{kt})-\R(c_{lr},c_{jr})-\R(c_{lr},c^{ij}_{st})-\R(c_{is},c_{js}).\label{RT2}\ee
We now calculate each of these terms in turn to obtain $\M$.

Using formulae from Appendix \ref{sect:sevenptrel}, we find the following:

(a) residue from $c_{lr}=c_{kt}=0$
\begin{align} \R(c_{lr},c_{kt})&={[i,j]\over(p^2)^2}\left[{\partial (c_{lr},c_{kt})\over \partial(\beta_k,\beta_l)}\right]^{-1} {c_{ir}c_{jt}\over\C_{kt}\C_{lt}} \label{Rcktclr}\\
&=-{1\over [i,j] W_t W_r c_{kr}c_{ks}c_{it}c_{ls}c_{lt}c_{jr}(c_{is}c_{jr}-c_{ir}c_{js})(c_{it}c_{js}-c_{is}c_{jt})}\nonumber\\
&=- {W_t^4W_r^4 [i,j]^3\over
\langle k,r\rangle\langle k,s\rangle \langle l,s\rangle\langle l,t\rangle
\;\langle s|P_{lt} |j]
\;\langle s | P_{kr} | i] \;
\langle t|P_{ij}P_{kr}|s\rangle\;
\langle s|P_{ij}P_{lt}|r\rangle}.
 \label{Rcktclr1}\end{align}

(b) residue from $c_{it}=c_{kt}=0$

Since at  $c_{it}=c_{kt}=0$,
\be {\partial (\C_{kt},c_{kt})\over \partial(\beta_k,\beta_l)}=-{ \partial\C_{kt}\over \partial\beta_l}{\partial c_{kt}\over\partial\beta_k}
=-{\partial c_{kt}\over\partial\beta_k}{ \partial c_{it}\over \partial\beta_l}c_{kr}c_{ks}c_{jt}(c_{is}c_{jr}-c_{ir}c_{js}),\nonumber\ee
\begin{align} \R(c_{it},c_{kt})
&={[i,j]\over(p^2)^2}\left[{\partial (\C_{kt},c_{kt})\over \partial(\beta_k,\beta_l)}\right]^{-1} {c_{ir}c_{jt}\over c_{lr}\C_{lt}} \label{Rcktcit}\\
&=-{1\over [j,l] W_t^2 c_{lr}c_{kr}c_{ks}c_{is}c_{jt}c_{lt}(c_{is}c_{jr}-c_{ir}c_{js})(c_{js}c_{lr}-c_{jr}c_{ls})}\nonumber\\
&={W_t^4 [j,l]^4\over
p^2_{ltj} \langle i,r\rangle \langle r,k\rangle \langle k,s\rangle\,
[l,t][t,j] \langle s|P_{lt}|j]\,\langle i|P_{jt} | l]}. \label{Rcktcit1}\end{align}
 
(c) residue from $c^{ij}_{rs}=c_{kt}=0$

Since at  $c^{ij}_{rs}=c_{kt}=0$,
\be {\partial (\C_{kt},c_{kt})\over \partial(\beta_k,\beta_l)}=-{ \partial\C_{kt}\over \partial\beta_l}{\partial c_{kt}\over\partial\beta_k}
={\partial c_{kt}\over\partial\beta_k}{ \partial c^{ij}_{rs}\over \partial\beta_l}c_{kr}c_{ks}c_{it}c_{jt},\nonumber\ee
\be{ \partial c^{ij}_{rs}\over \partial\beta_l}=[(c_{js}W_r-c_{jr}W_s)[j,l]+(c_{ir}W_s-c_{is} W_r)[l,i]]/p^2
=-\langle t|P_{ij}|l]/p^2\nonumber\ee
\be \C_{lt}|_{c^{ij}_{rs}=0}= c_{is}c_{lt}c^{ij}_{tr}c^{jl}_{rs}.\label{Clt01}\ee
\begin{align} \R(c^{ij}_{rs}, c_{kt}) 
&={[i,j]\over(p^2)^2}\left[{\partial (\C_{kt},c_{kt})\over \partial(\beta_k,\beta_l)}\right]^{-1} {c_{ir}c_{jt}\over c_{lr}\C_{lt}} \label{Rcktcijrs}\\
&=-{c_{ir}\over \langle t|P_{ij}|l] W_t c_{lr}c_{kr}c_{ks}c_{it}c_{is}c_{lt} (c_{jr}c_{it} - c_{jt}c_{ir}) (c_{jr}c_{ls} - c_{js}c_{lr})}\nonumber\\
&={W_t^4 \;\langle t|P_{ij}|l]^4\over
p^2_{ijt}p^2_{krs}  \langle r,k\rangle \langle k,s\rangle \langle i,j\rangle
\langle j,t\rangle\,
\langle r|P_{ks}|l]\,\langle i|P_{jt} | l]\;\langle t|P_{ij}P_{kr}|s\rangle}.\label{Rcktcijrs1}\end{align}

(d) residue from $c_{lr}=c_{jr}=0$

Since at  $c_{lr}=c_{jr}=0$,
\be {\partial (c_{lr},\C_{lt})\over \partial(\beta_k,\beta_l)}=-{ \partial c_{lr}\over \partial\beta_l}{\partial \C_{lt}\over\partial\beta_k}
={\partial c_{jr}\over\partial\beta_k}{ \partial c_{lr}\over \partial\beta_l}c_{ir}c_{ls}c_{lt}(c_{js}c_{it}-c_{is}c_{jt}),\nonumber\ee
\begin{align} \R(c_{lr},c_{jr})
&={[i,j]\over(p^2)^2}\left[{\partial (c_{lr},\C_{lt})\over \partial(\beta_k,\beta_l)}\right]^{-1} {c_{ir}c_{jt}\over \C_{kt}c_{kt}} \label{Rcjrclr}\\
&=-{1\over [k,i] W_r^2 c_{ls}c_{lt}c_{ir}c_{kr}c_{js}c_{kt}(c_{js}c_{it}-c_{is}c_{jt})(c_{is}c_{kt}-c_{it}c_{ks})}\nonumber\\
&= {W_r^4 [k,i]^4\over
p^2_{irk} \langle s,l\rangle \langle l,t\rangle \langle t,j\rangle\,
[i,r][r,k] \langle j|P_{ir}|k]\,\langle s|P_{kr}|i]}. \label{Rcjrclr1}\end{align}

(e) residue from $c_{lr}=c^{ij}_{st}=0$

Since at  $c_{lr}=c^{ij}_{st}=0$,
\be {\partial (c_{lr},\C_{lt})\over \partial(\beta_k,\beta_l)}=-{ \partial c_{lr}\over \partial\beta_l}{\partial \C_{lt}\over\partial\beta_k}
={\partial c^{ij}_{st}\over\partial\beta_k}{ \partial c_{lr}\over \partial\beta_l}c_{ls}c_{lt}c_{ir}c_{jr},\nonumber\ee
\be{ \partial c^{ij}_{st}\over \partial\beta_k}=[(c_{is}W_t-c_{it}W_s)[k,i]+(c_{jt}W_s-c_{js} W_t)[j,k]]/p^2
=-\langle r|P_{ij}|k]/p^2\nonumber\ee
\be \C_{kt}|_{c^{ij}_{st}=0}= c_{it}c_{kr}c^{jk}_{st}c^{ij}_{rs}.\label{Ckt01}\ee
\begin{align}  \R(c_{lr},c^{ij}_{st})
&={[i,j]\over(p^2)^2}\left[{\partial (c_{lr},\C_{lt})\over \partial(\beta_k,\beta_l)}\right]^{-1} {c_{ir}c_{jt}\over \C_{kt}c_{kt}}\label{Rcijstclr}\\
&=-{c_{jt}\over \langle r|P_{ij}|k] W_r c_{jr}c_{ls} c_{lt} c_{kt} c_{it} c_{kr}(c_{ir} c_{js} - c_{is} c_{jr}) (c_{js}c_{kt}- c_{ks}c_{jt})}\nonumber\\
&={W_r^4 \, \langle r|P_{ij}|k]^4\over
p^2_{ijr} p_{lst}^2\,\langle ji\rangle \langle ir\rangle
\langle sl\rangle \langle lt\rangle
\; \langle j|P{ir}| k ]
\;\langle t|P_{ls}|k]\; \langle s| P_{lt}P_{ij}|r\rangle}.\label{Rcijstclr1}\end{align}

(f) residue from $c_{is}=c_{js}=0$

Since at  $c_{is}=c_{js}=0$,
\be {\partial (\C_{kt},\C_{lt})\over \partial(\beta_k,\beta_l)}= {\partial (\C_{kt},\C_{lt})\over \partial(c_{is},c_{js})}\cdot {\partial (c_{is},c_{js})\over \partial(\beta_k,\beta_l)},\nonumber\ee
\begin{align}
 {\partial (\C_{kt},\C_{lt})\over \partial(c_{is},c_{js})}&=
c_{ir}c_{jr}c_{ks}c_{ls}c_{it}c_{jt}( c_{it}c_{jr}- c_{jt}c_{ir})( c_{kr}c_{lt}- c_{kt}c_{lr})
\cr
{\partial (c_{is},c_{js})\over \partial(\beta_k,\beta_l)}&= {[i,j][k,l]W_s^2\over (p^2)^2},
\nonumber\end{align}
\begin{align} \R(c_{is},c_{js})
&={[i,j]\over(p^2)^2}\left[{\partial (\C_{kt},\C_{lt})\over \partial(\beta_k,\beta_l)}\right]^{-1} {c_{ir}c_{jt}\over c_{lr}c_{kt}}\label{Rciscjs}\\
&={1\over [k,l]W_s^2 c_{jr}c_{lr}c_{ks}c_{ls}c_{it}c_{kt}( c_{it}c_{jr}- c_{jt}c_{ir})( c_{kr}c_{lt}- c_{kt}c_{lr})}\nonumber\\
&={W_s^4 [k,l]^4\over
p^2_{kls} \langle t,j\rangle \langle j,i\rangle \langle i,r\rangle\, 
[l,s][s,k] \langle r|P_{ks}|l]\,\langle t|P_{ls}| k]}.\label{Rciscjs1}\end{align}

Combining these results for the residues, we have that
\begin{align}
\M&=-\R(c_{lr},c_{kt})-\R(c_{it},c_{kt})-\R(c^{ij}_{rs},c_{kt})-\R(c_{lr},c_{jr})-\R(c_{lr},c^{ij}_{st})-\R(c_{is},c_{js})\cr
&={W_t^4W_r^4 [i,j]^3\over\langle k,r\rangle\langle k,s\rangle \langle l,s\rangle\langle l,t\rangle\langle s|P_{lt} |j]\langle s | P_{kr} | i] \langle t|P_{ij}P_{kr}|s\rangle\langle s|P_{ij}P_{lt}|r\rangle}\cr
&-{W_t^4 [j,l]^4\over p^2_{ltj} \langle i,r\rangle \langle r,k\rangle \langle k,s\rangle[l,t][t,j] \langle s|P_{lt}|j]\langle i|P_{jt} | l]}\cr
&-{W_t^4 \;\langle t|P_{ij}|l]^4\over p^2_{ijt}p^2_{krs}  \langle r,k\rangle \langle k,s\rangle \langle i,j\rangle\langle j,t\rangle\langle r|P_{ks}|l]\,\langle i|P_{jt} | l]\;\langle t|P_{ij}P_{kr}|s\rangle}\cr
&-{W_r^4 [k,i]^4\over p^2_{irk} \langle s,l\rangle \langle l,t\rangle \langle t,j\rangle[i,r][r,k] \langle j|P_{ir}|k]\,\langle s|P_{kr}|i]}\cr
&-{W_r^4 \, \langle r|P_{ij}|k]^4\over p^2_{ijr} p_{lst}^2\,\langle ji\rangle \langle ir\rangle\langle sl\rangle \langle lt\rangle\langle j|P{ir}| k ]\langle t|P_{ls}|k]\; \langle s| P_{lt}P_{ij}|r\rangle}\cr
&-{W_s^4 [k,l]^4\over p^2_{kls} \langle t,j\rangle \langle j,i\rangle \langle i,r\rangle[l,s][s,k] \langle r|P_{ks}|l]\,\langle t|P_{ls}| k]}.\label{MR2}
\end{align}

The expression (\ref{MR2}) 
has the appropriate soft limits and is 
antisymmetric under the transformations $i\leftrightarrow j$, 
$k\leftrightarrow l$, $r\leftrightarrow t$, $s\leftrightarrow s$. 
It is equal, up to notation and signs, to the expressions computed directly
from the field theory, and by recursion relations \cite{BDDK,BCF}.

Other tree amplitudes can be computed in a similar fashion from the expressions
for the integrand in section 4. It seems that the dual S-matrix of ACCK
leads back to twistor string theory at tree level. 
It will be interesting to pursue this link at the loop level.

\vss
\section*{Acknowledgements}
We are grateful to  Nima Arkani-Hamed for discussions, who also informed us that some similar results had been obtained by Marcus Spradlin and Anastasia Volovich \cite{SV}. LD thanks the Institute for 
Advanced Study at Princeton for its hospitality. 
LD was partially supported by the U.S. Department of Energy, 
Grant No. DE-FG02-06ER-4141801, Task A.

\appendix
\section{\bf Interchange of Particles between $\P$ and $\N$.}
\label{sect:Intpn}

For $k\in\P$, $t\in\N$, let $\P_0=\P\sim\{k\}$, $\N_0=\N\sim \{t\}$, and $\P'=\P_0\cup\{t\}$, $\N'=\N_0\cup\{k\}$,
\begin{align}
\pi_i &=\sum_{r\in\N_0} c_{ir}\pi_r+c_{it}\pi_t,\qquad i\in\P\cr
c_{kt}\pi_i-c_{it}\pi_k&=\sum_{r\in\N_0}(c_{ir}c_{kt}-c_{it}c_{kr})\pi_r, \qquad i\in\P_0\cr
\pi_i&=\sum_{r\in\N_0}{(c_{ir}c_{kt}-c_{it}c_{kr})\over c_{kt}} \pi_r+{c_{it}\over c_{kt}}\pi_k, \qquad i\in\P_0\cr
\pi_t&= -\sum_{r\in\N_0} {c_{kr}\over c_{kt}}\pi_r +{1\over c_{kt}}\pi_k.
\nonumber\end{align}
So
$$\pi_i=\sum_{r\in\N'}\tilde c_{ir}\pi_r,\qquad r\in\P',$$
with
$$\tilde c_{ir}={c_{ir}c_{kt}-c_{it}c_{kr}\over c_{kt}},\; i\in\P_0,r\in\N_0;
\quad
\tilde c_{ik}={c_{it}\over c_{kt}},\; i\in\P_0;\quad\tilde c_{tr}=-{c_{kr}\over c_{kt}},\; r\in\N_0;\quad
\tilde c_{tk}={1\over c_{kt}}.
$$
Similarly
$$-\bar\pi_r=\sum_{i\in\P'}\bar\pi_i\tilde c_{ir},\qquad r\in\N',$$
with the same definition of $\tilde c_{ir}$.
It remains to show that the $\tilde c_{ir}$ satisfy the same constraints as the $c_{ir}$. To do this we establish formulae for them in terms of the $\kappa_\ell, \rho_\ell$.
\begin{align}
\tilde c_{js}={c_{js}c_{kt}-c_{jt}c_{ks}\over c_{kt}}&={\kappa_j\over \kappa_s}
\left[\prod_{r\in\N\atop r\ne s}{\rho_j-\rho_r\over\rho_s-\rho_r}
-\prod_{r\in\N\atop r\ne t}{\rho_j-\rho_r\over\rho_k-\rho_r}
\prod_{r\in\N\atop u\ne s}{\rho_k-\rho_u\over\rho_s-\rho_u}\right],\qquad 
j\in\P_0,\;s\in\N_0,\cr
&={\kappa_j\over \kappa_s}\left[{\rho_j-\rho_t\over\rho_s-\rho_t}
\cdot{\rho_s-\rho_k
\over\rho_j-\rho_k}
-{\rho_j-\rho_s\over\rho_k-\rho_s}\cdot{\rho_k-\rho_t\over\rho_s-\rho_t}\cdot
{\rho_s-\rho_k\over\rho_j-\rho_k}\right]
\prod_{r\in\N'\atop r\ne s}{\rho_j-\rho_r\over\rho_s-\rho_r}\cr
&={\kappa_j\over \kappa_s}
\prod_{r\in\N'\atop r\ne s}{\rho_j-\rho_r\over\rho_s-\rho_r},\qquad j\in\P_0,
\;s\in\N_0.\nonumber\end{align}
$$
\tilde c_{jk}={c_{jt}\over c_{kt}}={\kappa_j\over \kappa_k}
\prod_{r\in\N\atop r\ne t}{\rho_j-\rho_r\over\rho_t-\rho_r}
\prod_{r\in\N\atop r\ne t}{\rho_t-\rho_r\over\rho_k-\rho_r}
={\kappa_j\over \kappa_k}\prod_{r\in\N'\atop r\ne k}{\rho_j-\rho_r\over\rho_k-\rho_r},\qquad j\in\P_0.
$$
\begin{align}
\tilde c_{ts}=-{c_{ks}\over c_{kt}}&=-{\kappa_t\over \kappa_s}
\prod_{r\in\N\atop r\ne s}
{\rho_k-\rho_r\over\rho_s-\rho_r}
\prod_{r\in\N\atop r\ne t}{\rho_t-\rho_r\over\rho_k-\rho_r}
={\kappa_t\over \kappa_s}\prod_{r\in\N'\atop r\ne s}
{\rho_t-\rho_r\over\rho_s-\rho_r}
\qquad s\in\N_0.
\nonumber\end{align}
$$
\tilde c_{tk}={1\over c_{kt}}={\kappa_t\over \kappa_k}\prod_{r\in\N\atop r\ne t}
{\rho_t-\rho_r\over\rho_k-\rho_r}
={\kappa_t\over \kappa_k}\prod_{r\in\N'\atop r\ne k}{\rho_t-\rho_r\over\rho_k-\rho_r}.
$$
Thus the $\tilde c_{js}$ are given by similar expressions in terms of the $\rho_\ell, k_\ell$ as the $c_{js}$, and so the $\tilde c_{js}$  satisfy similar relations to those satisfied by the $ c_{js}$.
\section{\bf Relations for the 6-point Function}
\label{sect:sixptrel}

$$V_jA_{ir}-V_iA_{jr}=-p^2[k,r],\qquad A_{ir}W_s-A_{is}W_r=-p^2\langle i,t\rangle$$
$$V_jA_{ir}W_s+V_iA_{js}W_r-V_jA_{is}W_r-V_iA_{jr}W_s=p^2\bA_{tk}$$
where
$$ \bA_{ri}=\sum_{ s \in\N}\langle r, s \rangle[ s ,i]=-\sum_{ j \in\P}\langle r, j \rangle[ j ,i], \qquad
\ba_{ri}={1\over p^2}\bA_{ri}$$

$$\det c={\beta\over p^2} \sum_{ i \in\P\atop r \in\N}V_ i  W_ r (a_{ j  s }a_{ k  t }-a_{ j  t }a_{ k  s })
={\beta\over (p^2)^2} \sum_{ i \in\P}V_ i  \bV_ i  \sum_{ r \in\N}W_ r \bW_ r =\beta,$$
where, as before, $( i , j , k )$, $( r , s , t )$ are cyclic. 

Corresponding to the relations for $A_{ir}$, we have
$$
\bA_{ri}\bA_{sj}-\bA_{si}\bA_{rj}= p^2W_tV_k,
$$
$$\bA_{ri} \bV_j-\bA_{rj}\bV_i=-p^2\langle r,k\rangle,$$
$$\bW_s \bA_{ri}-\bW_r\bA_{si}=-p^2[ t,i],$$
$$\bW_s\bA_{ri}\bV_j+\bW_r\bA_{sj}\bV_i-\bW_r\bA_{si}\bV_j-\bW_s\bA_{rj}\bV_i=p^2A_{kt}$$
$$
\bc_{ri}\bc_{sj}-\bc_{si}\bc_{rj}= \bbeta c_{kt}.
$$

When $c_{ir}=0$,
$$V_iW_r\beta\bc_{ri}=p^2_{jkr},$$
$$ V_iW_r\beta\bc_{sj}= \langle i,s\rangle [r,j],\quad V_iW_r\beta\bc_{tk}=\langle i,t\rangle [r,k],\quad
V_iW_r\beta\bc_{sk}= \langle i,s\rangle [r,k],\quad V_iW_r\beta\bc_{tj}=\langle i,t\rangle [r,j],$$
$$W_rc_{is}=\langle i,t\rangle,\quad W_rc_{it}=-\langle i,s\rangle,\quad V_ic_{jr}=[k,r],\quad V_ic_{kr}=-[j,r],$$
$$V_iW_r\beta\bc_{rj}= \langle i|P_{rk}|j] ,\quad V_iW_r\beta\bc_{rk}=\langle i|P_{rj}|k] ,
\quad V_iW_r\beta\bc_{si}=\langle s|P_{ti}|r],\quad V_iW_r\beta\bc_{ti}=\langle t|P_{si}|r],$$
$$V_iW_rc_{js}=[k|P_{si}|t\rangle,\quad V_iW_rc_{kt}=[j|P_{ti}|s\rangle,$$
using notation defined in (\ref{bracket}) and (\ref{defp}).

When $\bc_{ir}=0$,
$$\bV_i\bW_r\bbeta c_{ir}=p^2_{jkr},$$
$$ \bV_i\bW_r\bbeta c_{js}= [ i,s] \langle r,j\rangle ,\quad \bV_i\bW_r\bbeta c_{kt}=[ i,t] \langle r,k\rangle ,\quad
\bV_i\bW_r\bbeta c_{ks}= [ i,s] \langle r,k\rangle ,\quad \bV_i\bW_r\bbeta c_{jt}=[ i,t] \langle r,j\rangle ,$$
$$\bW_r\bc_{si}=-[ i,t],\quad \bW_r\bc_{ti}=[ i,s],\quad \bV_i\bc_{rj}=-\langle k,r\rangle ,\quad \bV_i\bc_{rk}=\langle j,r\rangle ,$$
$$\bV_i\bW_r\bbeta c_{jr}= [ i|P_{rk}|j\rangle  ,\quad \bV_i\bW_r\bbeta c_{kr}=[ i|P_{rj}|k\rangle  ,
\quad \bV_i\bW_r\bbeta c_{is}=[ s|P_{ti}|r\rangle ,\quad \bV_i\bW_r\bbeta c_{it}=[ t|P_{si}|r\rangle ,$$
$$\bV_i\bW_r\bc_{sj}=-\langle k|P_{si}|t],\quad \bV_i\bW_r\bc_{tk}=-\langle j|P_{ti}|s],$$
where $\bbeta=1/\beta$.

\section{\bf Relations for the 7-point Function}
\label{sect:sevenptrel}

We derive the relations we need to evaluate the 7-point function working directly from the equations
\begin{align}
\pi_\ell&=c_{\ell r}\pi_r+c_{\ell s}\pi_s+c_{\ell t}\pi_t,\qquad\qquad\quad \ell = i,j,k,l,\label{p1}\\
-\bpi_u&=\bpi_i c_{iu}+\bpi_j c_{ju}+\bpi_k c_{ku}+\bpi_l c_{lu},\qquad u=r,s,t.\label{p2}
\end{align}

If $c_{kt}=0$, from (\ref{p1}) with $\ell =k$,
\be
c_{kr}=\langle k,s\rangle/\langle r,s\rangle,\qquad c_{ks}=\langle k,r\rangle/\langle s,r\rangle,\qquad c_{kt}=0.\label{ckt}
\ee

(a) For $c_{kt}=c_{lr}=0$, in addition to (\ref{ckt}), 
\be c_{lr}=0,\qquad c_{ls}=\langle l,t\rangle/\langle s,t\rangle,\qquad c_{lt}=\langle l,s\rangle/\langle t,s\rangle.\label{cktclr1}
\ee
and from (\ref{p2}) with $u=r$,
$$\bpi_i c_{ir}+\bpi_j c_{jr}=-(\bpi_k \langle k,s\rangle+\bpi_r\langle r,s\rangle)/\langle r,s\rangle,$$
yielding
\be  c_{ir}=[j|P_{kr}|s\rangle/[i,j]\langle r,s\rangle,\qquad
c_{jr}=[i|P_{kr}|s\rangle/[j,i]\langle r,s\rangle,\label{cktclr2}\ee
and, similarly,
\be c_{it}=[j|P_{lt}|s\rangle/[i,j]\langle t,s\rangle,\qquad
c_{jt}=[i|P_{lt}|s\rangle/[j,i]\langle t,s\rangle.\label{cktclr3}\ee
From (\ref{p1}) with $\ell =i$
$$c_{is}\langle s,t\rangle =\langle i,t\rangle -c_{ir}\langle r,t\rangle,\qquad c_{js}\langle s,t\rangle =\langle j,t\rangle -c_{jr}\langle r,t\rangle$$
implying
$$(c_{ir}c_{js}-c_{jr}c_{is})\langle s,t\rangle=\langle j,t\rangle c_{ir}-\langle i,t\rangle c_{jr}=-\langle t|P_{ij}P_{kr}|s\rangle/[i,j]\langle r,s\rangle$$
so that 
\be
c_{ir}c_{js}-c_{jr}c_{is}=-\langle t|P_{ij}P_{kr}|s\rangle/[i,j]\langle r,s\rangle\langle s,t\rangle,\label{cktclr4}
\ee
and, similarly,
\be
c_{it}c_{js}-c_{jt}c_{is}=-\langle r|P_{ij}P_{lt}|s\rangle/[i,j]\langle s,t\rangle\langle r,s\rangle.\label{cktclr5}
\ee

(b) When $c_{kt}=c_{it}=0$, from (\ref{p1}) with $\ell =k,i$,
\be c_{kr}=\langle k,s\rangle/\langle r,s\rangle,\qquad c_{ks}=\langle k,r\rangle/\langle s,r\rangle,\qquad c_{kt}=0,\label{cktcit1}
\ee
and
\be
c_{ir}=\langle i,s\rangle/\langle r,s\rangle,\qquad c_{is}=\langle i,r\rangle/\langle s,r\rangle,\qquad c_{it}=0.\label{cktcit2}
\ee
From (\ref{p2}) with $u=t$,
\be
c_{jt}=-[t,l]/[j,l],\qquad c_{lt}=-[t,j]/[l,j],\label{cktcit3}\ee
and from (\ref{p1}) with $\ell=j,l$,
\be
c_{jr}\langle r,s\rangle=\langle s|P_{jt}|l]/[l,j],\qquad
c_{js}\langle s,r\rangle =\langle r|P_{jt}|l]/[l,j],\label{cktcit4}\ee
\be
c_{lr}\langle r,s\rangle=\langle s|P_{lt}|j]/[j,l],\qquad
c_{ls}\langle s,r\rangle =\langle r|P_{lt}|j]/[j,l].\label{cktcit5}\ee
From (\ref{p1}) with $\ell=i,j,$
$$\langle r,s\rangle^2[l,j](c_{ir}c_{js}-c_{is}c_{jr})=\langle i,r\rangle\langle s|P_{jt}|l]-\langle i,s\rangle \langle r|P_{jt}|l] =-\langle r,s\rangle \langle i|P_{jt}|l]$$
so that
\be
c_{ir}c_{js}-c_{is}c_{jr}= \langle i|P_{jt}|l]/\langle r,s\rangle[j,l]\label{cktcit6}
\ee
From (\ref{p1}) with $\ell=j,l,$
$$
(c_{lr}c_{js}-c_{jr}c_{ls})\langle s,t\rangle=\langle j,t\rangle c_{lr}-\langle l,t\rangle c_{jr}=(\langle s|P_{lt}|j]\langle j,t\rangle +\langle s|P_{jt}|l]\langle l,t\rangle) /[j,l]\langle r,s\rangle,
$$
from which it follows that
\be
c_{lr}c_{js}-c_{jr}c_{ls}=-p_{jlt}^2/[j,l]\langle r,s\rangle.\label{cktcit7}
\ee

(c) When $c_{kt}=c_{rs}^{ij}=0$, from (\ref{p1}) with $\ell=k$
\be
c_{kr}=\langle k,s\rangle/\langle r,s\rangle,\qquad c_{ks}=\langle k,r\rangle/\langle s,r\rangle,\qquad c_{kt}=0,
\label{cktcrsij1}\ee
 and with $\ell =i,j$,
$$c_{jr}\pi_i-c_{ir}\pi_j=(c_{jr}c_{it}-c_{ir}c_{jt})\pi_t,$$
so that
\be
c_{jr}\langle i,t\rangle=c_{ir}\langle j,t\rangle, \qquad c_{js}\langle i,t\rangle=c_{is}\langle j,t\rangle.\ee
From (\ref{p2}) with $u=r$,
$$-[l,r]=[l,i] c_{ir}+[l,j] c_{jr}+[l,k] \langle k,s\rangle/\langle r,s\rangle,$$
implying
$$-\langle s|P_{kr}|l]/\langle r,s\rangle=[l,i] c_{ir}+[l,j] c_{jr}=\langle t|P_{ij}|l]c_{ir}/\langle i,t\rangle,$$
so that
\be
c_{ir}=-{\langle s|P_{kr}|l]\langle i,t\rangle\over\langle t|P_{ij}|l]\langle r,s\rangle},\quad c_{jr}=-{\langle s|P_{kr}|l]\langle j,t\rangle\over\langle t|P_{ij}|l]\langle r,s\rangle},
\quad c_{jr}c_{it}-c_{ir}c_{jt}={\langle s|P_{kr}|l]\langle i,j\rangle\over\langle t|P_{ij}|l]\langle r,s\rangle},\label{cktcrsij2}\ee
\be
c_{is}=-{\langle r|P_{ks}|l]\langle i,t\rangle\over\langle t|P_{ij}|l]\langle s,r\rangle},\quad c_{js}=-{\langle r|P_{ks}|l]\langle j,t\rangle\over\langle t|P_{ij}|l]\langle s,r\rangle},
\quad c_{js}c_{it}-c_{is}c_{jt}={\langle r|P_{ks}|l]\langle i,j\rangle\over\langle t|P_{ij}|l]\langle s,r\rangle},\label{cktcrsij3}\ee
and, from (\ref{p1}) with $\ell=i$,
$$c_{it}\langle t|P_{ij}|l]\langle r,s\rangle\langle i,t\rangle
=-\langle s|P_{kr}|l]\langle i,t\rangle\langle r,i\rangle+\langle r|P_{ks}|l]\langle i,t\rangle\langle s,i\rangle
=-\langle r,s\rangle \langle i|P_{rks}|l]\langle i,t\rangle$$
so that
\be
c_{it}={ \langle i|P_{jt}|l]\over \langle t|P_{ij}|l]},\qquad c_{jt}={ \langle j|P_{it}|l]\over \langle t|P_{ij}|l]}.\label{cktcrsij4}\ee
From (\ref{p2}) with $u=t$,
$$\langle t|P_{ij}|l][t,l]c_{lt}=-[t,i]c_{it}-[t,j]c_{jt}=-[t,i]\langle i|P_{jt}|l]-[t,j]\langle j|P_{it}|l]$$
implying
\be c_{lt}={ p^2_{ijt}\over \langle t|P_{ij}|l]}.\label{cktcrsij5}\ee
From (\ref{p1}) with $\ell=l$,
$$ \langle t|P_{ij}|l]c_{lr}\langle r,s\rangle=\langle l,s\rangle \langle t|P_{ij}|l] -c_{lt}\langle t,s\rangle \langle t|P_{ij}|l]=\langle l,s\rangle \langle t|P_{ij}|l] -p^2_{ijt}\langle t,s\rangle
=- \langle t|P_{ij}P_{kr}|s\rangle$$
so that
\be
c_{lr}=-{\langle t|P_{ij}P_{kr}|s\rangle\over \langle r,s\rangle  \langle t|P_{ij}|l]},\qquad c_{ls}=-{\langle t|P_{ij}P_{ks}|r\rangle\over \langle s,r\rangle  \langle t|P_{ij}|l]}.\label{cktcrsij6}\ee
From (\ref{p1}) with $\ell=j,l$,
$$(c_{jr}c_{ls}-c_{js}c_{lr})\langle r,t\rangle=c_{ls}\langle j,t\rangle-c_{js}\langle l,t\rangle$$
implying
\be c_{jr}c_{ls}-c_{js}c_{lr}=-{p_{krs}^2\langle j,t\rangle\over \langle r,s\rangle  \langle t|P_{ij}|l]}.\label{cktcrsij7}\ee

(d) When $c_{lr}=c_{jr}=0$, proceeding as in (b), we have
\be
c_{lr}=0,\qquad c_{ls}=\langle l,t\rangle/\langle s,t\rangle,\qquad c_{lt}=\langle l,s\rangle/\langle t,s\rangle,\label{clrcjr1}
\ee
\be
c_{jr}=0,\qquad c_{js}=\langle j,t\rangle/\langle s,t\rangle,\qquad c_{jt}=\langle j,s\rangle/\langle t,s\rangle,\label{clrcjr2}\ee
\be
c_{ir}=-[r,k]/[i,k],\qquad c_{kr}=-[r,i]/[k,i],\label{clrcjr3}\ee
\be
c_{is}\langle s,t\rangle=\langle t|P_{ir}|k]/[k,i],\qquad c_{it}\langle t,s\rangle =\langle s|P_{ir}|k]/[k,i],\label{clrcjr4}\ee\
\be
c_{ks}\langle s,t\rangle=\langle t|P_{kr}|i]/[i,k],\qquad c_{kt}\langle t,s\rangle =\langle s|P_{kr}|i]/[i,k],\label{clrcjr5}\ee
\be
c_{it}c_{js}-c_{is}c_{jt}=\langle j|P_{ir}|k]/\langle s,t\rangle[i,k],\label{clrcjr6}\ee
\be
c_{is}c_{kt}-c_{it}c_{ks}=p_{ikr}^2/[i,k]\langle s,t\rangle.\label{clrcjr7}\ee

(e) When $c_{lr}=c_{st}^{ij}=0$, proceeding as in (c), we have
\be
c_{lr}=0,\qquad c_{ls}=\langle l,t\rangle/\langle s,t\rangle,\qquad c_{lt}=\langle l,s\rangle/\langle t,s\rangle,\label{clrcstij1}\ee
\be
c_{it}=-{\langle s|P_{lt}|k]\langle i,r\rangle\over\langle r|P_{ij}|k]\langle t,s\rangle},\qquad c_{jt}=-{\langle s|P_{lt}|k]\langle j,r\rangle\over\langle r|P_{ij}|k]\langle t,s\rangle},
\qquad c_{jt}c_{ir}-c_{it}c_{jr}={\langle s|P_{lt}|k]\langle i,j\rangle\over\langle r|P_{ij}|k]\langle t,s\rangle},\label{clrcstij2}\ee
\be
c_{is}=-{\langle t|P_{ls}|k]\langle i,r\rangle\over\langle r|P_{ij}|k]\langle s,t\rangle},\qquad c_{js}=-{\langle t|P_{ls}|k]\langle j,r\rangle\over\langle r|P_{ij}|k]\langle s,t\rangle},
\qquad c_{js}c_{ir}-c_{is}c_{jr}={\langle t|P_{ls}|k]\langle i,j\rangle\over\langle r|P_{ij}|k]\langle s,t\rangle},\label{clrcstij3}\ee
\be
c_{ir}={ \langle i|P_{jr}|k]\over \langle r|P_{ij}|k]},\qquad c_{jr}={ \langle j|P_{ir}|k]\over \langle r|P_{ij}|k]},\label{clrcstij4}\ee
\be
c_{kr}={ p^2_{ijr}\over \langle r|P_{ij}|k]},\label{clrcstij5}\ee
\be
c_{kt}=-{\langle r|P_{ij}P_{lt}|s\rangle\over \langle t,s\rangle  \langle r|P_{ij}|k]},\qquad c_{ks}=-{\langle r|P_{ij}P_{ls}|t\rangle\over \langle s,t\rangle  \langle r|P_{ij}|k]},\label{clrcstij6}\ee
\be
c_{jt}c_{ks}-c_{js}c_{kt}=-{p_{lst}^2\langle j,r\rangle\over \langle t,s\rangle  \langle r|P_{ij}|k]}.\label{clrcstij7}\ee

(f) When  $c_{is}=c_{js}=0$, also proceeding as in (b),
we have
\be
c_{ir}=\langle i,t\rangle/\langle r,t\rangle,\qquad c_{is}=0,\qquad c_{it}=\langle i,r\rangle/\langle t,r\rangle,\label{ciscjs1}\ee
\be
c_{jr}=\langle j,t\rangle/\langle r,t\rangle,\qquad c_{js}=0,\qquad c_{jt}=\langle j,r\rangle/\langle t,r\rangle,\label{ciscjs2}\ee
\be
c_{ks}=-[s,l]/[k,l],\qquad c_{ls}=-[s,k]/[l,k],\label{ciscjs3}\ee
\be
c_{kr}\langle r,t\rangle=\langle t|P_{ks}|l]/[l,k],\qquad c_{kt}\langle t,r\rangle =\langle r|P_{ks}|l]/[l,k],\label{ciscjs4}\ee
\be
c_{lr}\langle r,t\rangle=\langle t|P_{ls}|k]/[k,l],\qquad c_{lt}\langle t,r\rangle =\langle r|P_{ls}|k]/[k,l],\label{ciscjs5}\ee
\be
c_{it}c_{jr}-c_{ir}c_{jt}=-(\langle i,r\rangle\langle j,t\rangle-\langle j,r\rangle\langle i,t\rangle)/\langle r,t\rangle^2=\langle i,j\rangle/\langle t,r\rangle.\label{ciscjs6}\ee
\be
(c_{kr}c_{lt}-c_{lr}c_{kt})\langle r,s\rangle=\langle k,s\rangle c_{lt}-\langle l,s\rangle c_{kt}=(\langle r|P_{ls}|k]\langle k,s\rangle +\langle r|P_{ks}|l]\langle l,s\rangle) /[k,l]\langle t,r\rangle,
\nonumber\ee
so that
\be
c_{kr}c_{lt}-c_{lr}c_{kt}=-p_{kls}^2/[k,l]\langle t,r\rangle.\label{ciscjs7}\ee

\vfill\eject

\singlespacing


\providecommand{\bysame}{\leavevmode\hbox to3em{\hrulefill}\thinspace}
\providecommand{\MR}{\relax\ifhmode\unskip\space\fi MR }
\providecommand{\MRhref}[2]{%
  \href{http://www.ams.org/mathscinet-getitem?mr=#1}{#2}
}
\providecommand{\href}[2]{#2}

\end{document}